\documentclass[11pt,a4paper]{article}
\pdfoutput=1

\usepackage{jheppub,dsfont}
\usepackage{amsmath}
\usepackage{slashed}

\def\ba{\begin{eqnarray}}
\def\ea{\end{eqnarray}}

\def\be{\begin{equation}}
\def\ee{\end{equation}}

\newcommand{\checked}[1]{}

\title{The static hard-loop gluon propagator to all orders in anisotropy}

\author[a]{Mohammad Nopoush,}
\affiliation[a]{Department of Physics, Kent State University, Kent, OH 44242 United States}

\author[b]{Yun Guo,}
\affiliation[b]{Department of Physics, Guangxi Normal University, Guilin, 541004, China}

\author[a]{and Michael Strickland\hspace{0.25mm}}

\abstract{We calculate the (semi-)static hard-loop self-energy and propagator using the Keldysh formalism in a momentum-space anisotropic quark-gluon plasma.  The static retarded, advanced, and Feynman (symmetric) self-energies and propagators are calculated to all orders in the momentum-space anisotropy parameter $\xi$.  For the retarded and advanced self-energies/propagators, we present a concise derivation and comparison with previously-obtained results and extend the calculation of the self-energies to next-to-leading order in the gluon energy, $\omega$.  For the Feynman self-energy/propagator, we present new results which are accurate to all orders in $\xi$.  We compare our exact results with prior expressions for the Feynman self-energy/propagator which were obtained using Taylor-expansions around an isotropic state.  We show that, unlike the Taylor-expanded results, the all-orders expression for the Feynman propagator is free from infrared singularities.  Finally, we discuss the application of our results to the calculation of the imaginary-part of the heavy-quark potential in an anisotropic quark-gluon plasma.}

\begin{document} 
\maketitle
\flushbottom

\section{Introduction}
\label{sec:intro}

When nuclear matter is heated to extremely high temperatures at low net baryon density, it experiences a crossover from ordinary hadronic matter to a new phase in which quark and gluon quasiparticles are the relevant degrees of freedom.  The high-temperature phase is called the quark-gluon plasma (QGP) and in order to study it experimentally, physicists use ultrarelativistic heavy-ion collisions (URHICs) performed using, for example, the Relativistic Heavy Ion Collider at Brookhaven National Laboratory and the Large Hadron Collider at the European Organization for Nuclear Research.  Due to the short lifetime of the QGP created in URHICs ($\tau \lesssim 15$ fm/c), in order to study the behavior of the QGP,  external probes, such as ultrafast laser pulses or relativistic particle beams, are infeasible. Instead, one must rely on internal probes which are created as a part of system and are subject to the system's full spatiotemporal evolution.  If one is interested in probing the full dynamics of the QGP, however, light hadrons which are formed at late times during hadronization do not provide a complete picture.  To complement studies of light hadron production, heavy-ion theorists and experimentalists also study particles that are produced at early times and, ideally, do not interact significantly in the final state medium, e.g. dilepton pairs and heavy quarkonia states.  Heavy quarkonia production in URHICs, in particular, can provide information about the initial temperature of the QGP, deviations from equilibrium, etc., due to the fact that bound states of heavy quarks are formed very early in the collision and do not melt until the temperature is a few times the phase transition temperature \cite{Mocsy:2013syh,Andronic:2015wma}.  In order to apply this idea to phenomenology, however, one needs to compute the in-medium spectrum of bound states.  This step is traditionally done in the context of a finite-temperature equilibrium quantum field theory and the result of this exercise is then folded together with a model for the spatiotemporal evolution of the QGP.  However, in practice this approach is insufficient since, at early times, the QGP  possesses a significant degree of momentum-space anisotropy.

To address the deviations from isotropic equilibrium, one could attempt to compute the non-equilibrium corrections to the heavy quark potential ``perturbatively'' as is done in the framework of viscous hydrodynamics, however, at very early times after the collision or near the transverse/longitudinal edges of the QGP this method will become unreliable since such approaches rely on an implicit assumption that the system is near equilibrium.  In order to address this problem, one can reorganize the hydrodynamic expansion around a state that is anisotropic in momentum-space at leading-order.  This reorganized expansion has been dubbed the ``anisotropic hydrodynamics'' framework and it is now being widely applied to heavy-ion phenomenology, see e.g. Refs.~\cite{Strickland:2014pga,Florkowski:2010cf,Martinez:2010sc,Martinez:2012tu,Ryblewski:2012rr,Bazow:2013ifa,Tinti:2013vba,Nopoush:2014pfa,Tinti:2015xwa,Bazow:2015cha,Strickland:2015utc,Alqahtani:2015qja,Molnar:2016vvu,Molnar:2016gwq,Alqahtani:2016rth,Alqahtani:2017jwl} and references therein.  In this approach one assumes a specific form for the momentum-space anisotropic distributions function for the light quarks and gluons which is of Romatschke-Strickland (RS) form \cite{Romatschke:2003ms,Romatschke:2004jh} or generalized RS form \cite{Tinti:2013vba,Nopoush:2014pfa,Alqahtani:2015qja,Kasmaei:2016apv}.  Since momentum-space anisotropies affect e.g. the binding energies and widths of heavy-quarkonia states, there has been a concerted effort to compute their effect on the heavy-quark potential itself \cite{Dumitru:2007hy,Dumitru:2009ni,Burnier:2009yu,Dumitru:2009fy,Du:2016wdx,Biondini:2017qjh} which, when convoluted with a realistic anisotropic hydrodynamics background evolution, has resulted in phenomenological calculations of the nuclear suppression of bottomonia which are in good agreement with available experimental data \cite{Strickland:2011mw,Strickland:2011aa,Krouppa:2015yoa,Krouppa:2016jcl,Krouppa:2017lsw}.  

 One of the shortcomings of all previous calculations of the heavy-quark potential in a momentum-space anisotropic QGP is that, although the real part of the potential was obtained to all orders in the plasma momentum-space anisotropy parameter $\xi$ in Ref.~\cite{Dumitru:2007hy}, calculations of the imaginary part of the heavy-quark potential have relied on a Taylor expansion around $\xi=0$ to linear order in $\xi$~\cite{Burnier:2009yu}. In practice, the in-medium heavy-quark potential is obtained from the Fourier transform of the sum of the static dressed retarded, advanced, and Feynman propagators, i.e.
\be
V({\bf r},\xi) = - \frac{g^2 C_F}{2} \int \frac{d^3{\bf p}}{(2\pi)^3} \left(e^{i {\bf p}\cdot {\bf r}} - 1 \right) 
\left[ \tilde{\cal D}^{00}_R + \tilde{\cal D}^{00}_A + \tilde{\cal D}^{00}_F \right]_{\omega \rightarrow 0} ,
\ee
\checked{mn}
where $g$ is the strong coupling constant, $C_F = (N_c^2-1)/(2 N_c)$ is the quadratic Casimir in the fundamental representation of $SU(N_c)$.  The first two terms involving the retarded and advanced propagators are purely real and, consequently, the imaginary part of the potential comes solely from the Fourier transform of the Feynman propagator.  As a result, in order to compute the imaginary part of the potential, one must first compute the dressed Feynman propagator to all orders in $\xi$.  In this paper, we perform this computation, laying the groundwork for the all-orders calculation of the imaginary part of the heavy-quark potential in a momentum-space anisotropic QGP.  In the process of the derivation, we repeat some previously obtained results for completeness of the presentation, e.g. we revisit the computation of the exact expressions for the retarded and advanced propagators obtained originally in Ref.~\cite{Dumitru:2007hy}.  In order to carry out the calculation, we make use of the real-time Keldysh formalism and we work in the high-temperature ``hard-loop'' limit~\cite{Carrington:1997sq}.  As a cross-check, our final results for the static limit of the hard-loop Feynman propagator are compared to previously-obtained Taylor-expanded results.

\section{Setup and notation}
\label{sec:seutp}
In this paper, the metric is taken to be ``mostly minus'', i.e. in Minkowski space $\eta^{\mu\nu}\equiv(1,-\mathds{1})$. Lower-case letters denote four-vectors and bold lower-case letters denote three vectors. All Greek-letter indices stand for the components of four-vectors while Latin indices indicate spatial components of four-vectors. For any two four-vectors $x^\mu\equiv (x_0,{\bf x})$ and $y^\mu\equiv (y_0,{\bf y})$ the inner product is defined as $x\cdot y\equiv x^\mu y_\mu=x_0 y_0-{\bf x}\cdot{\bf y}$. The subscripts `R', `A', and `F' for propagators and self-energies stand for retarded, advanced, and Feynman propagators, respectively. According to Feynman slash notation, for any four-vector $x$, we have $\slashed{x} \equiv \gamma\cdot x$, with $\gamma^\mu$ being Dirac matrices. Herein, the anisotropic distribution function is taken to be of Romatschke-Strickland (RS) form \cite{Romatschke:2003ms,Romatschke:2004jh}
\be
f_{\rm aniso}({\bf p})\equiv f_{\rm iso}\!\left(\frac{1}{\lambda}\sqrt{{\bf p}^2+\xi ({\bf p}\cdot {\bf n})^2}\right) ,
\ee
\checked{m}
where ${\bf n}$ is a three-dimensional unit vector, $\lambda$ is a temperature-like scale which corresponds to temperature in the equilibrium limit, $\xi$ is the momentum-space anisotropy parameter which, for $-1<\xi<0$, corresponds to a prolate momentum-space anisotropy and, for $\xi>0$, corresponds to an oblate momentum-space anisotropy. Herein, we assume that the chemical potential is zero. $f_{\rm iso}$ is an equilibrium distribution function, i.e. $f_{\rm iso}(x)\equiv [\exp(x)+a]^{-1}$, which depending on the particles statistics can be Boltzmann ($a=0$), Fermi-Dirac $f_F$, ($a=1$), or Bose-Einstein $f_B$, ($a=-1$) distribution functions
\be
f_{F\!/\!B}({\bf p})=\left[\exp\!\left(\frac{1}{\lambda}\sqrt{{\bf p}^2+\xi ({\bf p\cdot n})^2}\right)\pm 1\right]^{-1}\,. 
\label{eq:FB-dist}
\ee
\checked{mn}
 We will perform the calculation initially in QED and then, in the end, we will generalize our results to QCD.

\section{The real-time formalism for non-equilibrium field theories}

In this section, we present the basic formalism used to obtain our results in a concise and self-contained manner.  We will use the real-time Keldysh formalism \cite{Dumitru:2009fy, Carrington:1998jj, Mrowczynski:2000ed, Mrowczynski:2016etf}. The Keldysh formalism is based on contour Green's functions. For a spinor field $\psi$ and a vector field $A^\mu$, we can define the fermionic and bosonic QED Green's functions as the following, respectively
\ba
i(S(x,y))_{\alpha\beta}\equiv \langle \hat{T}[\psi_\alpha(x)\bar{\psi}_\beta(y)]\rangle \,,\\
i({\cal D}(x,y))_{\mu\nu}\equiv \langle \hat{T}[A_\mu(x) A_\nu(y)]\rangle\,,
\label{eq:gfuncs1}
\ea
\checked{mn}
where $\{\alpha,\beta\} \in \{1,2,3,4\}$ are spinor indices, $\{\mu,\nu\} \in \{0,1,2,3\}$ are Lorentz indices. The angle brackets $\langle\cdots\rangle$ denote the quantum expectation value and $\hat{T}$ is the time-ordering operator defined as 
\be 
\hat{T}[X(x)Y(y)]\equiv \Theta(x_0-y_0)X(x)Y(y)\pm \Theta(y_0-x_0)Y(y)X(x)\,,
\label{eq:timeorder1}
\ee
\checked{mn}
where $\Theta$ is the Heaviside step function. Corresponding to different ways of propagation with respect to the contour, one can define four functions based on the contour propagator:
\ba 
i(S^>(x,y))_{\alpha\beta} &\equiv&  \langle \psi_\alpha(x)\bar{\psi}_\beta(y)\rangle\,, \nonumber \\
i(S^<(x,y))_{\alpha\beta}&\equiv& -\langle \bar{\psi}_\beta(y)\psi_\alpha(x)\rangle\,, \nonumber \\
i(S^c(x,y))_{\alpha\beta}&\equiv& \langle \hat{T}^c[\psi_\alpha(x)\bar{\psi}_\beta(y)]\rangle\,, \nonumber \\
i(S^a(x,y))_{\alpha\beta}&\equiv& \langle \hat{T}^a[\psi_\alpha(x)\bar{\psi}_\beta(y)]\rangle\,. 
\label{eq:gfuncs2}
\ea
\checked{mn}
Likewise, for bosons, we have 
\ba 
i({\cal D}^>(x,y))_{\mu\nu} &\equiv&  \langle A_\mu(x)A_\nu(y)\rangle\,, \nonumber \\
i({\cal D}^<(x,y))_{\mu\nu}&\equiv& \langle A_\nu(y)A_\mu(x)\rangle\,, \nonumber \\
i({\cal D}^c(x,y))_{\mu\nu}&\equiv& \langle \hat{T}^c[A_\mu(x)A_\nu(y)]\rangle\,, \nonumber \\
i({\cal D}^a(x,y))_{\mu\nu}&\equiv& \langle \hat{T}^a[A_\mu(x)A_\nu(y)]\rangle\,. 
\label{eq:gfuncs3}
\ea
\checked{mn}
 In the relation above, $\hat{T}^c$ and $\hat{T}^a$ are time-ordering and anti-time-ordering operators, respectively, which are defined as
\ba 
\hat{T}^c[X(x)Y(y)]&\equiv& \Theta(x_0-y_0)X(x)Y(y)\pm \Theta(y_0-x_0)Y(y)X(x) \,,\\
\hat{T}^a[X(x)Y(y)]&\equiv& \Theta(y_0-x_0)X(x)Y(y)\pm \Theta(x_0-y_0)Y(y)X(x)\,.
\label{eq:timeorder2}
\ea
\checked{mn}
The upper and lower signs above correspond to bosonic and fermionic cases, respectively. In practice, one can define four different Green's functions with the following meanings:
\ba
S^c(x,y)&\equiv& S(x,y)\quad \text{\small with both $x_0$ and $y_0$ on the upper branch}\,, \nonumber \\
S^a(x,y)&\equiv& S(x,y)\quad \text{\small with both $x_0$ and $y_0$ on the lower branch}\,, \nonumber  \\
S^<(x,y)&\equiv& S(x,y)\quad \text{\small with $x_0$ on the upper and $y_0$ on the lower branch}\,, \nonumber  \\
S^>(x,y)&\equiv& S(x,y)\quad \text{\small with $x_0$ on the lower and $y_0$ on the upper branch}\,.\nonumber 
\label{eq:gfunction3}
\ea
\checked{m}
In the Keldysh formulation it is useful to introduce the following $2\times 2$ matrix 
\ba
S=
  \begin{pmatrix}
     S_{11} & S_{12} \\ S_{21} & S_{22}
     \end{pmatrix} 
     =  \begin{pmatrix}
     S^c & S^< \\ S^> & S^a
     \end{pmatrix}\,.
     \label{eq:keldysh}
\ea
\checked{m}
Since, the system under consideration is assumed to be translationally invariant, the two-point function only depends on $x-y$. Therefore, we can safely set $y=0$ and study the two point function function of a single variable ($x$). The components of the electron propagator $S$ satisfy the following relations
\ba
S^{c/a}(x)&=&\Theta(x_0)S^\gtrless(x)+\Theta(-x_0) S^\lessgtr(x)\,, \\
S^c(x)+S^a(x)&=&S^<(x)+S^>(x) \,.
\ea
\checked{m}
The retarded, advanced, and Feynman propagators are defined as 
\ba
S_{R/A} (x) &\equiv& \pm \Theta(\pm x_0)(S^>(x)-S^<(x))\,,\\
S_{R/A} (-x) &=& \pm \Theta(\mp x_0)(S^>(-x)-S^<(-x))\,,\\
S_F(x)&=& S^>(x)+S^<(x)\,.
\label{eq:SF<>}
\ea
\checked{m}
Using the above relations one finds some useful identities such as
\ba
S_R(\pm x)-S_A(\pm x)&=&S^>(\pm x)-S^<(\pm x)\,, \label{eq:SRA<>}\\
 S_{R/A}(x)&=&\pm \Theta(\pm x_0)[S_R(x)-S_A(x)]\,,
 \label{eq:SRA} \\
  S_{R/A}(-x)&=&\pm \Theta(\mp x_0)[S_R(- x)-S_A(- x)]\,.
 \label{eq:SRA2}
\ea
\checked{mn}
The one-loop photon self-energy is defined as
\be
\Pi^{\mu\nu}(x)=-ie^2 {\rm Tr}[\gamma^\mu S(x) \gamma^\nu S(-x)]\,.
\ee
\checked{m}
The retarded, advanced, and Feynman self-energies are defined as following
\ba
\Pi_{R/A}^{\mu\nu}(x)&=&\pm \Theta(\pm x_0)(\Pi^>(x)-\Pi^<(x))\,,
\label{eq:PiRA<>}\\
\Pi_F^{\mu\nu}(x)&=&\Pi^>(x)+\Pi^<(x)\,,
\label{eq:PiF<>}
\ea
\checked{m}
with
\be
\left(\Pi^\lessgtr(x)\right)^{\mu\nu} =-i e^2{\rm Tr} [\gamma^\mu S^\lessgtr(x)\gamma^\nu S^\gtrless(-x)]\,.
\label{eq:Pi<>}
\ee
\checked{m}
Substituting (\ref{eq:Pi<>}) into (\ref{eq:PiRA<>}), the retarded/advanced self-energy can be written as
\be
\Pi_{R/A}^{\mu\nu}(x)= \mp ie^2 \Theta(\pm x_0) {\rm Tr}[\gamma^\mu S^>(x) \gamma^\nu S^<(-x)-\gamma^\mu S^<(x) \gamma^\nu S^>(-x)]\,.
\label{eq:PiR<>}
\ee
\checked{mn}
Using (\ref{eq:SF<>}) and (\ref{eq:SRA<>}) we have
\ba
S^>(x)=\frac{S_R(x)-S_A(x)+S_F(x)}{2}\,, \nonumber \\
S^<(x)=\frac{-S_R(x)+S_A(x)+S_F(x)}{2}\,.
\label{eq:S<>}
\ea
\checked{m}
Thus, Eq.~(\ref{eq:PiR<>}) gives
\ba
\Pi_{R/A}^{\mu\nu}(x) &=& \mp \frac{i}{2}e^2 \Theta(\pm x_0) {\rm Tr}[\gamma^\mu S_F(x) \gamma^\nu S_A(-x)-\gamma^\mu S_F(x) \gamma^\nu S_R(-x)\nonumber \\
&& \hspace{3cm}
+\gamma^\mu S_R(x) \gamma^\nu S_F(-x)-\gamma^\mu S_A(x) \gamma^\nu S_F(-x)] \nonumber \\
&=& -\frac{i}{2}e^2 {\rm Tr}[\gamma^\mu S_F(x) \gamma^\nu S_{A/R}(-x)
+\gamma^\mu S_{R/A}(x) \gamma^\nu S_F(-x)]\,.
\ea
\checked{mn}
where in the last line we have used (\ref{eq:SRA}). Performing the Fourier transform of both sides  we have
\be
\Pi_{R/A}^{\mu\nu}(p)=-i\frac{e^2}{2}\int \frac{d^4k}{(2\pi)^4}{\rm Tr}[\gamma^\mu S_{R/A}(k)\gamma^\nu S_F(q)+\gamma^\mu S_F(k)\gamma^\nu S_{A/R}(q)]\,,
\label{eq:PiR0}
\ee
\checked{mn}
with $q\equiv k-p$. The Feynman self-energy can be obtained by substituting (\ref{eq:Pi<>}) in (\ref{eq:PiF<>}) and then using (\ref{eq:S<>})
\ba
 \Pi_F^{\mu\nu}(x) &=&-ie^2{\rm Tr}[\gamma^\mu S^>(x)\gamma^\nu S^<(-x)+\gamma^\mu S^<(x)\gamma^\nu S^>(-x)]\nonumber \\ 
 &=&  -i\frac{e^2}{2}{\rm Tr}\Big[\gamma^\mu S_F(x)\gamma^\nu S_F(-x)-\gamma^\mu \left[S_R(x)-S_A(x)\right]\gamma^\nu [S_R(-x)-S_A(-x)]\Big]\,.
\ea
\checked{mn}
After performing the Fourier transform of both sides one has
\be 
 \Pi_F^{\mu\nu}(p) =  -i\frac{e^2}{2}\int \frac{d^4k}{(2\pi)^4}{\rm Tr}  \Big[\gamma^\mu S_F(k)\gamma^\nu S_F(q)-\gamma^\mu \left[S_R(k)-S_A(k)\right]\gamma^\nu \left[S_R(q)-S_A(q)\right]\Big]\,.
 \label{eq:PiF0}
\ee
\checked{m}
We will calculate the hard loop self-energies and propagators using the Keldysh real-time formalism in the limit of vanishing chemical potential~\cite{Carrington:1997sq}. The  ``bare'' propagators are then $2 \times 2 $ matrices such as
\ba
S(k)= &&  \left (\begin{array}{cc}
\frac{\slashed{k}}{k^2+i\epsilon} & 0\\
0 & \frac{-\slashed{k}}{k^2-i\epsilon}\\
                          \end{array} \right )
 +2\pi i\, \slashed{k}\,\delta (k^2)\>\left (\begin{array}{cc}
f_F({\bf k}) & -\Theta (-k_0)+f_F({\bf k})\\
-\Theta (k_0)+f_F({\bf k}) & f_F({\bf k}) \\ \end{array} \right )\,,
\label{2a5}
\ea
\checked{m}
for a massless Dirac field and 
\ba
 \label{2a4}
  {\cal D}(k)  =  \left (\begin{array}{cc} \frac{1}{k^2+i\epsilon} & 0\\
                             0 & \frac{-1}{k^2-i\epsilon}\\
            \end{array} \right ) -  2\pi i\, \delta (k^2)
 \left (\begin{array}{cc}
f_B({\bf k}) & \Theta (-k_0)+f_B({\bf k})\\
\Theta (k_0)+f_B({\bf k}) & f_B({\bf k}) \\ \end{array} \right )\,,
\ea
\checked{m}
for a massless scalar field. In the above relations, $\epsilon$ is a small positive number which is sent to zero only at the end of calculation. It should be noted that since Eqs.~(\ref{2a5}) and (\ref{2a4}) are bare propagators, the hard-loop resummation has yet to be performed. The retarded, advanced, and Feynman propagators can be obtained from the Keldysh representation (which satisfies ${\cal D}_{11}-{\cal D}_{12}-{\cal D}_{21}+{\cal D}_{22}=0$) via 
\ba
\label{2a6}
   {\cal D}_R = {\cal D}_{11} - {\cal D}_{12} ~,~ {\cal D}_A = {\cal D}_{11} - {\cal D}_{21} ~,~
   {\cal D}_F = {\cal D}_{11} + {\cal D}_{22}  ~,
\ea
\checked{m}
with analogous expressions holding for the fermionic propagators. In momentum space, the explicit expressions for the bare propagators as a function of a general momentum $k$ are
\ba
S_R(k) & = & \frac{\slashed{k}}{k^2+i\, \mbox{sgn}(k_0) \epsilon},\nonumber \\
S_A(k) & = & \frac{\slashed{k}}{k^2-i\, \mbox{sgn}(k_0) \epsilon},\nonumber \\
S_F(k) & = & -2\pi i\, \slashed{k} \, [1-2f_F({\bf k})]\, \delta (k^2)\,,
\label{2a11}
\ea
\checked{m}
for fermions and 
\ba
{\cal D}_R(k) & = & \frac{1}{k^2+i\, \mbox{sgn}(k_0) \epsilon},\nonumber \\
{\cal D}_A(k) & = & \frac{1}{k^2-i\, \mbox{sgn}(k_0) \epsilon},\nonumber \\
{\cal D}_F(k) & = & -2\pi i\, [1+2f_B({\bf k})]\, \delta (k^2)\,, \label{2a10}
\ea
\checked{m}
for scalar bosons. 
In the real-time formalism, the following relations hold for the self
energies:
\ba
\Pi_{11}+\Pi_{12}+\Pi_{21}+\Pi_{22}=0 \label{2a12}\,,
\ea
\checked{m}
and
\ba
\Pi_R  =  \Pi_{11}+\Pi_{12} ~,~
\Pi_A  =  \Pi_{11}+\Pi_{21} ~,~
\Pi_F  =  \Pi_{11}+\Pi_{22} ~. \label{2a13}
\ea
\checked{m}
Note that, for vector fields, one must add the appropriate Lorentz indices to the propagators and self-energies.

\subsection*{The dressed photon propagator}
The dressed photon propagator can be obtained from the Dyson-Schwinger equation 
\be
i\tilde{\cal D}=i{\cal D}+i{\cal D}(-i\Pi)i\tilde{\cal D} \,,
\ee
\checked{m}
where, in the Keldysh formalism, both the propagators and self-energies are $2\times2$ matrices, ${\cal D}$ is the bare photon propagator and $\tilde{\cal D}$ is the dressed propagator. For the dressed retarded propagator, we have 
\be 
\tilde{\cal D}_{R/A}={\cal D}_{R/A}+{\cal D}_{R/A}\Pi_{R/A} \tilde{\cal D}_{R/A}\,.
\label{eq:ds-PiR}
\ee
\checked{m}
The dressed Feynman propagator satisfies
\be
\tilde{\cal D}_F={\cal D}_{F}+{\cal D}_{R}\Pi_R \tilde{\cal D}_F+{\cal D}_{F}\Pi_A \tilde{\cal D}_A+{\cal D}_{R}\Pi_F \tilde{\cal D}_A \,,
\ee 
\checked{m}
which, upon using $p = (\omega,{\bf p})$ and ${\cal D}_F(p)=[1+2f_B({\bf p})]{\rm sgn}(\omega)[{\cal D}_R(p)-{\cal D}_A(p)]$, becomes
\ba
\tilde{\cal D}_{F}(p)= && (1+2f_B({\bf p}))\, \mbox{sgn}(\omega)\,
[\tilde{\cal D}_{R}(p)-\tilde{\cal D}_{A}(p)]
\nonumber \\
&& +\tilde{\cal D}_{R}(p)\,\{\Pi _F(p)-(1+2f_B({\bf p}))\, \mbox{sgn}(\omega)\, [\Pi
_R(p)-\Pi _A(p)]\} \,  \tilde{\cal D}_A(p)~. \label{eq:2b8}
\ea
\checked{m}
Note that in the relations introduced so far the distribution functions are general. We will specify the precise forms in the forthcoming sections.

\section{Tensor decomposition in a momentum-space anisotropic plasma }

Since the photon propagator is a tensor, we must find a suitable tensor basis and extract the corresponding scalar coefficient functions.  For anisotropic systems there are more independent projectors than for the standard equilibrium case due to the fact that there is an additional spacelike vector ${\bf n}$ which defines the anisotropy direction~\cite{Romatschke:2003ms}. Here, we use a four-tensor basis which is appropriate for systems with one anisotropy direction~\cite{Dumitru:2007hy}. Specifically, we introduce four tensors
\ba
A^{\mu \nu}&=& -\eta^{\mu\nu}+\frac{p^\mu
p^\nu}{p^2}+\frac{\tilde{m}^\mu \tilde{m}^\nu}{\tilde{m}^2}\,,\nonumber \\
B^{\mu \nu}&=& -\frac{p^2}{(m\cdot p) ^2}\frac{\tilde{m}^\mu
\tilde{m}^\nu}{\tilde{m}^2}\,,\nonumber \\
C^{\mu \nu}&=& \frac{\tilde{m}^2p^2}{\tilde{m}^2p^2+(n\cdot
p)^2}[\tilde{n}^\mu
\tilde{n}^\nu-\frac{\tilde{m}\cdot\tilde{n}}{\tilde{m}^2}(\tilde{m}^\mu
\tilde{n}^\nu+\tilde{m}^\nu
\tilde{n}^\mu)+\frac{(\tilde{m}\cdot\tilde{n})^2}{\tilde{m}^4}\tilde{m}^\mu
\tilde{m}^\nu]\,,\nonumber \\
D^{\mu \nu}&=& \frac{p^2}{m\cdot p}
\left[ 2\frac{\tilde{m}\cdot\tilde{n}}{\tilde{m}^2}\tilde{m}^\mu
\tilde{m}^\nu-
\left(\tilde{n}^\mu \tilde{m}^\nu+\tilde{m}^\mu
\tilde{n}^\nu\right) \right]\,.
\label{eq:ABCD}
\ea
\checked{m}
Here, $m^{\mu}$ is the heat-bath four-velocity, which in the local rest frame is given by $m^{\mu}=(1,0,0,0)$, and
\begin{equation}
\tilde{m}^\mu=m^{\mu}-\frac{m\cdot p}{p^2} \,p^\mu\,,
\label{eq:mt}
\end{equation}
\checked{m}
is the component of $m^\mu$ orthogonal to $p^\mu$. The direction of anisotropy in momentum space is determined by the vector
\begin{equation}
n^{\mu}=(0,{\bf n})\,,
\label{eq:n}
\end{equation}
\checked{m}
where ${\bf n}$ is a three-dimensional unit vector. Likewise, $\tilde{n}^\mu$ is the component of $n^\mu$ orthogonal to $p^\mu$.
The self-energies and dressed propagators can be expanded in terms of the tensor basis as
\ba
\Pi^{\mu\nu}_{R,A,F} &=&\alpha_{R,A,F} A^{\mu\nu}+\beta_{R,A,F} B^{\mu\nu} + \gamma_{R,A,F}
C^{\mu\nu} + \delta_{R,A,F} D^{\mu\nu}\,,\label{eq:Piexp}\\
\tilde{\cal D}^{\mu\nu}_{R,A,F}&=&\alpha'_{R,A,F} A^{\mu\nu}+\beta'_{R,A,F} B^{\mu\nu} + \gamma'_{R,A,F}
C^{\mu\nu} + \delta'_{R,A,F} D^{\mu\nu}\,.
\label{eq:Dexp}
\ea
\checked{m}
We remind the reader that, due to the transversality of the photon self-energy $p_\mu \Pi^{\mu\nu}=0$, not all components of $\Pi^{\mu\nu}$ are independent.  One has four equations which can be used, for example, to write the timelike rows/columns of the self-energy tensor in terms of the space-like components
\ba 
\nu=0 \; &\Rightarrow& \;  \omega \Pi^{00}+p_i \Pi^{i0}=0\,,\\
\nu=i \; &\Rightarrow& \; \omega \Pi^{0i}+p_j \Pi^{ji}=0\,.
\ea
\checked{m}
Using the symmetry of $\Pi^{\mu\nu} = \Pi^{\nu\mu}$, one finds \mbox{$\omega^2\Pi^{00}=p_i\Pi^{ij}p_j$}. These relations show that having $\Pi^{ii}$ and $\Pi^{xy}$, $\Pi^{xz}$, $\Pi^{yz}$ (6 components overall), one can obtain all components of $\Pi^{\mu\nu}$. 

Restricting our attention to the spatial block of $\Pi^{\mu\nu}$, $\Pi^{ij}$, one can obtain the expansion tensor coefficients of self-energy using the following projections
\ba
p^i \Pi^{ij}p^j &=&{\bf p}^2 \beta\,, \nonumber \\
A^{il}n^l \Pi^{ij}p^j &=&({\bf p}^2-(n\cdot p)^2)\delta\,, \nonumber \\
A^{il}n^l \Pi^{ij}A^{jk}n^k &=&\frac{{\bf p}^2-(n\cdot p)^2}{{\bf p}^2}(\alpha+\gamma)\,, \nonumber \\
{\rm Tr}\,\Pi^{ij} &=&2\alpha+\beta+\gamma\,.
\label{eq:abgd}
\ea
\checked{m}
An alternative method for extracting the coefficient functions, which is based on the four-tensor form of $\Pi^{\mu\nu}$, is presented in Appendix~{\ref{app:alternative}}.

The dressed retarded/advanced propagators satisfy (\ref{eq:ds-PiR}), which can be solved to give
\be
\tilde{\cal D}_{R,A}^{-1}= ({\cal D}_{R,A})^{-1}-\Pi_{R,A}\,.
\ee
\checked{m}
Using the definition of bare propagator and (\ref{eq:Piexp}) one has
\ba
(\tilde{\cal D}_{R,A}^{-1})^{\mu\nu}&=&-p^2 \eta^{\mu\nu} +p^\mu p^\nu-\Pi_{R,A}^{\mu\nu}-\frac{1}{\zeta}p^\mu p^\nu \\
&=&(p^2-\alpha_{R,A})A^{\mu\nu}+(\omega^2-\beta_{R,A})B^{\mu\nu}-\gamma_{R,A} C^{\mu\nu}-\delta_{R,A} D^{\mu\nu}-\frac{1}{\zeta}p^\mu p^\nu\,.
\ea
\checked{mn}
where $\zeta$ is the gauge fixing parameter.  One can obtain $\tilde{\cal D}$ by inverting the above relation~\citep{Dumitru:2007hy}
\be
\tilde{\cal D}_{R,A}^{\mu\nu}\!=\!\Delta_A\left[A^{\mu\nu}-C^{\mu\nu}\right]\!+\!\Delta_G\left[(p^2-\alpha_{R,A}-\gamma_{R,A})\frac{\omega^4}{p^4}B^{\mu\nu}+(\omega^2-\beta_{R,A})C^{\mu\nu}+\delta_{R,A}\frac{\omega^2}{p^2}D^{\mu\nu}\right]\!-\!\frac{\zeta}{p^4}p^\mu p^\nu ,
\label{eq:tildeDRA}
\ee
\checked{m}
with 
\ba
\Delta_A^{-1} &=& p^2 - \alpha_{R,A} \label{eq:deltaa} \,,\\
\Delta_G^{-1} &=& (p^2-\alpha_{R,A}-\gamma_{R,A})(\omega^2-\beta_{R,A})-\delta_{R,A}^2\left[{\bf p}^2-(n\cdot p)^2\right]\,.
\label{eq:deltag}
\ea
\checked{mn}
Comparing to Eq.~(\ref{eq:Dexp}), the expansion coefficients are
\ba
\alpha'_{R,A}&=&\Delta_A\,,\nonumber\\
\beta'_{R,A}&=&\Delta_G(p^2-\alpha_{R,A}-\gamma_{R,A})\frac{\omega^4}{p^4}\,,\nonumber\\
\gamma'_{R,A}&=&\Delta_G(\omega^2-\beta_{R,A})-\Delta_A\,,\nonumber\\
\delta' _{R,A}&=&\Delta_G \frac{\omega^2}{p^2} \delta_{R,A}\,.
\label{eq:coefPrime}
\ea
\checked{mn}
Herein, we take the gauge parameter to be zero, $\zeta=0$, which is allowed since the static limit of the gauge propagator is gauge invariant.

\section{The retarded and advanced gluon self-energies}
%
In this section, we calculate the gluon self-energy in the hard-loop (HL) limit. Since, in the HL-limit, the photon and gluon self-energies are the same up to definition of Debye mass, we start with Eq.~(\ref{eq:PiR0}) for photon self-energy. Note that regarding Eq.~(\ref{eq:PiR0}) and (\ref{2a11}), $\Pi_R$ and $\Pi_A$ are complex conjugates of each other and we only need to find one of them. To proceed, we start with the retarded photon self-energy
\be
\Pi_R^{\mu\nu}(p,\xi)=-i\frac{e^2}{2}\int \frac{d^4k}{(2\pi)^4}{\rm Tr}[\gamma^\mu S_R(k)\gamma^\nu S_F(q)+\gamma^\mu S_F(k)\gamma^\nu S_A(q)]\,,
\ee
\checked{m}
where, specializing to anisotropic distribution function (\ref{eq:FB-dist}) in this section, we made the dependence on the anisotropy parameter, $\xi$,  explicit. Using $S_{R,A,F}(k)=\slashed{k}\Delta_{R,A,F}(k)$ with
\ba
\Delta_R(k) & = & \frac{1}{k^2+i\, \mbox{sgn}(k_0) \epsilon},\nonumber \\
\Delta_A(k) & = & \frac{1}{k^2-i\, \mbox{sgn}(k_0) \epsilon},\nonumber \\
\Delta_F(k) & = & -2\pi i\, \big[1-2f_F({\bf k})\big]\, \delta (k^2)\,,
\label{eq:propag}
\ea
\checked{m}
and ${\rm Tr}[\gamma^\mu\gamma^\alpha\gamma^\nu\gamma^\beta]=
4(\eta^{\mu\alpha}\eta^{\nu\beta}-\eta^{\mu\nu}\eta^{\alpha\beta}+\eta^{\mu\beta}\eta^{\alpha\nu})$, we have
\ba
\Pi^{\mu\nu}_R(p,\xi)&=&-2ie^2\int \frac{d^4k}{(2\pi)^4}(q^\mu k^\nu+ q^\nu k^\mu-\eta^{\mu\nu}q\cdot k) \Big[\Delta_R(k)\Delta_F(q)+\Delta_F(k)\Delta_A(q)\Big]\nonumber \\
&=&-4ie^2\int \frac{d^4k}{(2\pi)^4}(q^\mu k^\nu+ q^\nu k^\mu-\eta^{\mu\nu}q\cdot k)\Delta_F(k)\Delta_A(q)\,,
\label{eq:PiR2}
\ea
\checked{m}
where, in going from the first to the second line, we have used the fact that two terms in the integrand are equal under the transformation $k\rightarrow -k+p\,(=-q)$.
Using (\ref{eq:propag}), one has
\ba
  \Pi^{\mu\nu}_R(p,\xi)&=&16\pi e^2 \int \frac{d^4k}{(2\pi )^4}
f_F({\bf k}) \left[
\frac{q^\mu k^\nu+ q^\nu k^\mu-\eta^{\mu\nu}q\cdot k}{k^2+p^2-2k \cdot p-i\epsilon\,{\rm sgn}(k_0-\omega)}
\right]
\delta(k^2)\,.
\label{eq:PiR3}
\ea
\checked{m}
Note that the first term in $\Delta_F(k)$ in Eq.~(\ref{eq:propag}) corresponds to a divergent vacuum contribution.  This contribution is subtracted to obtain the in-medium photon self-energy. Now we can take the HL approximation, that is, taking all internal momenta $k$ to be of order $\lambda$ (hard) and the external momenta $p$ to be of order $e\lambda$ (soft) and Taylor-expand the integrand around $e=0$. At the leading order of the HL approximation, the quantity in square brackets in (\ref{eq:PiR3}) is
\be
\left[ \; \cdots \; \right] \longrightarrow \frac{2k^\mu k^\nu}{-2 k\cdot p-i \epsilon\,{\rm sgn}(k_0)}\,.
\ee
\checked{m}
Note that the terms containing $k^2$ are effectively zero due to the delta function which enforces the mass shell condition. Substituting this into the integral (\ref{eq:PiR3}), using 
\be
\delta(k^2)=\delta(k_0^2-{\bf k}^2)=\frac{1}{2|{\bf k}|} \Big(\delta(k_0-|{\bf k}|)+\delta(k_0+|{\bf k}|)\Big)\,,
\label{eq:delta-f}
\ee
\checked{m}
integrating over $k_0$, and finally setting ${\bf k}\rightarrow -{\bf k}$ in negative-energy contribution, one finds
\be
4e^2\int \frac{d^3{\bf k}}{(2\pi )^3}
\frac{f_F({\bf k})}{|{\bf k}|}
\bigg[\frac{-2k^\mu k^\nu}{2|{\bf k}|\omega-2{\bf k\cdot p}+i \epsilon\,}+\frac{2k^\mu k^\nu}{2|{\bf k}|\omega-2{\bf k\cdot p}+i \epsilon\,}\bigg]
\Bigg|_{k_0 = |{\bf k}|}
=0\,.
\ee
\checked{mn}
At next-to-leading order we have 
\be
\left[ \; \cdots \; \right] \longrightarrow  -\frac{\eta^{\mu\nu}}{2}+\frac{k^\mu p^\nu+k^\nu p^\mu}{2k\cdot p+i\epsilon\,{\rm sgn}(k_0)}-\frac{2k^\mu k^\nu p^2}{(2k\cdot p+i\epsilon\, {\rm sgn}(k_0))^2}\,,
\ee
where by substituting into the integral one obtains the retarded photon self-energy in the HL limit 
\ba
  \Pi^{\mu\nu}_R(p,\xi)&=&8\pi e^2 \int \frac{d^4k}{(2\pi )^4}
f_F({\bf k})
\bigg[-\eta^{\mu\nu}+\frac{k^\mu p^\nu+k^\nu p^\mu}{k\cdot p+i\epsilon\, {\rm sgn}(k_0)\,}-\frac{k^\mu k^\nu p^2}{(k\cdot p+i\epsilon\, {\rm sgn}(k_0))^2}\bigg]\delta(k^2)\,.
\label{eq:PiR4}
\ea
Performing the integral over $k_0$ and again setting ${\bf k}\rightarrow -{\bf k}$ in the negative energy contribution, one has
\be
  \Pi^{\mu\nu}_R(p,\xi)=4 e^2 \int \frac{d^3{\bf k}}{(2\pi )^3}
\frac{f_F({\bf k})}{|{\bf k}|}
\bigg[-\eta^{\mu\nu}+\frac{k^\mu p^\nu + k^\nu p^\mu}{ k\cdot p+i\epsilon}-\frac{k^\mu k^\nu p^2}{(k\cdot p+i\epsilon)^2}\bigg] \Bigg|_{k_0 = |{\bf k}|}
\,.
\label{eq:PiR5}
\ee
One can show that, for on-shell momentum $k^\mu$, 
\be
 |{\bf k}|\frac{\partial}{\partial k^l}\bigg[\frac{k^\mu k^\nu p^l}{|{\bf k}|(p\cdot k+i \epsilon)}-\frac{k^\mu \eta^{\nu l}}{|{\bf k}|}\bigg]=\frac{k^\mu p^\nu+k^\nu p^\mu}{p\cdot k+i \epsilon}-\frac{k^\mu k^\nu p^2}{(p\cdot k+i\epsilon)^2}-\eta^{\mu\nu}  \,,
\ee
so, we after integrating by parts one obtains
\ba
  \Pi^{\mu\nu}_R(p,\xi)&=&- 4 e^2 \int \frac{d^3{\bf k}}{(2\pi )^3}
\frac{\partial f_F({\bf k})}{\partial k^l} 
\bigg[\frac{k^\mu k^\nu p^l}{|{\bf k}|(p\cdot k+i \epsilon)}-\frac{k^\mu \eta^{\nu l}}{|{\bf k}|}\bigg] \Bigg|_{k_0 = |{\bf k}|}
\,. \label{eq:PiR6} 
\ea
\checked{m}
By specializing to the RS form for the anisotropic distribution function, we can simplify $\Pi_R^{\mu\nu}$ a bit more. 
Using the anisotropic Fermi-Dirac distribution (\ref{eq:FB-dist}), one has
\ba
\frac{\partial f_F({\bf k})}{\partial k^l} &=&\frac{v^l+\xi ({\bf v\cdot n}) n^l}{1+\xi({\bf v\cdot n})^2}\frac{\partial f_F({\bf k})}{\partial |{\bf k}|}\,, \label{eq:dist-identity1} \\
\frac{2e^2}{\pi^2} \int d|{\bf k}|\,{\bf k}^2\frac{\partial f_F({\bf k})}{\partial|{\bf k}|}&=&  \frac{1}{1+\xi({\bf v\cdot n})^2} \frac{2 e^2}{\pi^2}  \int d|{\bf k}|\,{\bf k}^2\frac{\partial f_F^{\rm iso}({\bf k})}{\partial|{\bf k}|}=-\frac{m_D^2}{1+\xi({\bf v\cdot n})^2}\,,
\label{eq:dist-identity2}
\ea
\checked{m}
where $v^\mu\equiv k^\mu/|{\bf k}|=(1,{\bf k}/|{\bf k}|)$ and the QED Debye mass is defined as
\be 
m_D^2\equiv -\frac{2e^2}{\pi^2}\int d|{\bf k}|{\bf k}^2 \frac{\partial f_F^{\rm iso}({\bf k})}{\partial |{\bf k}|}=\frac{e^2\lambda^2}{3}\,.
\label{eq:mDQED}
\ee
\checked{mn}
Substituting (\ref{eq:dist-identity1}) and (\ref{eq:dist-identity2}) into (\ref{eq:PiR6}), one obtains
\be
  \Pi^{\mu\nu}_R(p,\xi)= m_D^2 \int \frac{d\Omega}{4\pi} \,
v^\mu \frac{v^l+\xi ({\bf v\cdot n}) n^l}{(1+\xi({\bf v\cdot n})^2)^2} \bigg[-\eta^{\nu l} +\frac{v^\nu  p^l}{p\cdot v+i\epsilon}\bigg]\,.
\label{eq:PiR8}
\ee
\checked{m}
This precisely corresponds to the result obtained in Ref.~\cite{Mrowczynski:2000ed,Romatschke:2003ms} using relativistic kinetic theory.

\section*{The retarded and advanced hard-loop gluon self-energies}
\label{sec:retarded-PiR}

As mentioned previously, in the hard-loop limit, the photon and gluon self-energies are the same up to the definition of the Debye mass. The effective QCD distribution function includes contributions from quarks, anti-quarks, and gluons, including the degeneracy factors for the number of quark flavors and gluon color-charge states
\be
f({\bf k})=2N_c f_g({\bf k})+N_f (f_q({\bf k}) +f_{\bar q}({\bf k}))\,,
\label{eq:qcd-dist}
\ee
\checked{m}
where $f_g({\bf k})$ is gluonic distribution function, $f_q({\bf k})$ and $f_{\bar q}({\bf k})$ are quarks and anti-quarks distribution functions.
Eq.~(\ref{eq:PiR8}) is valid for the gluon self-energy  provided that QED Debye mass is replaced with its QCD counterpart, defined as
\be 
m_D^2\equiv -\frac{g^2}{2\pi^2}\int d|{\bf k}|{\bf k}^2 \frac{\partial f_{\rm iso}({\bf k})}{\partial |{\bf k}|}=\frac{(2N_c+N_f)g^2\lambda^2}{6}\,.
\label{eq:qcd-debye}
\ee
\checked{m}
 In order to calculate the expansion coefficients (\ref{eq:Piexp}) we only need the spatial block of $\Pi_R^{\mu\nu}$ which is
\be
  \Pi^{ij}_R(p,\xi)= m_D^2 \int \frac{d\Omega}{4\pi} \,
v^i \frac{v^l+\xi ({\bf v\cdot n}) n^l}{(1+\xi({\bf v\cdot n})^2)^2} \Bigg[\delta^{lj} +\frac{v^j  p^l}{p\cdot v+i\epsilon}\Bigg] .
\label{eq:PiRij}
\ee
\checked{m}
In order to obtain the tensor expansion coefficients for the retarded self-energy $\Pi_R^{\mu\nu}$ (\ref{eq:Piexp}) we can make use of Eqs.~(\ref{eq:abgd}). For this purpose, we need to choose a frame to be able to define the vectors ${\bf k}$, ${\bf p}$, and ${\bf n}$, and subsequently, the tensor basis matrices. The trivial choice is to take ${\bf n}$ along $z$-axis and ${\bf p}$ in the $x$-$z$ plane making an angle $\theta_n\equiv \arctan(p_x/p_z)$ with $z$-axis. Based on this coordinate we have 
\ba
{\bf n}&=&(0,0,1);\:\quad {\bf v}=(\sin\theta\cos\phi,\sin\theta\sin\phi,\cos\theta);\:\quad \hat{\bf p}=\left(\frac{p_x}{|{\bf p}|},0,\frac{p_z}{|{\bf p}|}\right)\,.
\label{eq:def1}
\ea
\checked{m}
Up to next-to-leading order in $\omega$, we have
\ba
\alpha_{R/A}(p,\xi)&=&-\frac{m_D^2}{2p_x^2} \Bigg[\frac{p_z^2\arctan\sqrt{\xi}}{\sqrt{\xi}}-\frac{{\bf p}^2 p_z\arctan\Big[\frac{p_z\sqrt{\xi}}{\sqrt{{\bf p}^2+\xi p_x^2}}\Big]}{\sqrt{\xi}\sqrt{{\bf p}^2+\xi p_x^2}}\pm i\frac{{\bf p}^2p_x^2 \pi(1+\xi)}{2({\bf p}^2+\xi p_x^2)^{3/2}}\omega\Bigg]+{\cal O}(\omega^2)\,,\nonumber \\
\beta_{R/A}(p,\xi)&=&-\frac{m_D^2\omega^2}{2{\bf p}^2}\Bigg[\frac{{\bf p}^2}{{\bf p}^2+\xi p_x^2}+\frac{\arctan\sqrt{\xi}}{\sqrt{\xi}}+\frac{{\bf p}^2 p_z\sqrt{\xi}\arctan\Big[\frac{p_z\sqrt{\xi}}{\sqrt{{\bf p}^2+\xi p_x^2}}\Big]}{({\bf p}^2+\xi p_x^2)^{3/2}}\nonumber \\&& \hspace{40mm}\pm i\frac{2{\bf p}^4+{\bf p}^2(2{\bf p}^2+p_x^2)\xi+p_x^2(p_x^2-p_z^2)\xi^2}{2({\bf p}^2+\xi p_x^2)^{5/2}}\pi\omega\Bigg]+{\cal O}(\omega^4)\,, \nonumber \\
\gamma_{R/A}(p,\xi)&=&-\frac{m_D^2}{2}\Bigg[\frac{{\bf p}^2}{{\bf p}^2+\xi p_x^2}-\frac{({\bf p}^2+p_z^2)\arctan\sqrt{\xi}}{\sqrt{\xi}p_x^2}+\frac{{\bf p}^2 p_z(2{\bf p}^2+3\xi p_x^2)\arctan\Big[\frac{p_z\sqrt{\xi}}{\sqrt{{\bf p}^2+\xi p_x^2}}\Big]}{p_x^2\sqrt{\xi}({\bf p}^2+\xi p_x^2)^{3/2}}\nonumber \\ &&\hspace{63mm}\mp i\frac{\xi\pi p_x^2({\bf p}^2+\xi({\bf p}^2+2p_z^2))}{2({\bf p}^2+\xi p_x^2)^{5/2}}\omega\Bigg]+{\cal O}(\omega^2)\,,\nonumber \\
\delta_{R/A}(p,\xi)&=&\frac{m_D^2\omega}{2{\bf p}^2p_x^2}\Bigg[\mp i \frac{{\bf p}^2p_x^2p_z\pi\xi}{2({\bf p}^2+\xi p_x^2)^{3/2}}+\frac{3{\bf p}^2p_x^2p_z\xi\omega}{({\bf p}^2+\xi p_x^2)^2}+\frac{p_z\omega}{\sqrt{\xi}}\arctan\sqrt{\xi}\nonumber \\ &&\hspace{25mm}- \frac{{\bf p}^2({\bf p}^4+\xi {\bf p}^2p_x^2-3\xi^2p_x^2p_z^2)\omega}{\sqrt{\xi}({\bf p}^2+\xi p_x^2)^{5/2}}\arctan\bigg[\frac{p_z\sqrt{\xi}}{\sqrt{{\bf p}^2+\xi p_x^2}}\bigg]\Bigg]+{\cal O}(\omega^3)\,. \hspace{1cm}
\label{eq:PiRcoef}
\ea
\checked{y}

\subsubsection*{Static limit}

One finds that, in the static limit, $\Delta_A$ (\ref{eq:deltaa}) and $\Delta_G$ (\ref{eq:deltag}) become
\ba
\lim_{\omega\rightarrow 0}\Delta_A^{-1} &=& -( {\bf p}^2 + m_\alpha^2) \, , \\
\lim_{\omega\rightarrow 0}\Delta_G^{-1} &=& -\frac{\omega^2}{{\bf p}^2}\left[({\bf p}^2+m_\alpha^2+m_\gamma^2)({\bf p}^2+m_\beta^2)
  - m_\delta^4 \right] . 
\ea
\checked{mn}
with \cite{Romatschke:2003ms,Dumitru:2007hy}
\ba
m_\alpha^2&\equiv&-\frac{m_D^2}{2 p_x^2 \sqrt{\xi}}%
\left(p_z^2 {\rm{arctan}}{\sqrt{\xi}}-\frac{p_z {\bf{p}}^2}{\sqrt{{\bf{p}}^2+\xi p_x^2}}%
{\rm{arctan}}\bigg[\frac{\sqrt{\xi} p_{z}}{\sqrt{{\bf{p}}^2+\xi p_x^2}}\bigg]\right) \; , \label{eq:malpha}\\
m_\beta^2&\equiv&m_{D}^2
\frac{(\sqrt{\xi}+(1+\xi){\rm{arctan}}{\sqrt{\xi}})({\bf{p}}^2+\xi p_x^2)+\xi p_z\left(%
p_z \sqrt{\xi} + \frac{{\bf{p}}^2(1+\xi)}{\sqrt{{\bf{p}}^2+\xi p_x^2}} %
{\rm{arctan}}\Big[\frac{\sqrt{\xi} p_{z}}{\sqrt{{\bf{p}}^2+\xi p_x^2}}\Big]\right)}{%
2  \sqrt{\xi} (1+\xi) ({\bf{p}}^2+ \xi p_x^2)}\,, \;\;\;\; \label{eq:mbeta} \\
m_\gamma^2&\equiv&-\frac{m_D^2}{2}\left(\frac{{\bf{p}}^2}{{\bf{p}}^2 + \xi p_x ^2}%
-\frac{{\bf p}^2+p_z^2}{\sqrt{\xi}p_x^2}{\rm{arctan}}{\sqrt{\xi}}+\frac{
p_z{\bf{p}}^2(2{\bf{p}}^2+3\xi p_x^2)}{\sqrt{\xi}(\xi
p_x^2+{\bf{p}}^2)^{{3/2}}
p_x^2}{\rm{arctan}}\bigg[\frac{\sqrt{\xi}
p_{z}}{\sqrt{{\bf{p}}^2+\xi p_x^2}}\bigg]\right) , \label{eq:mgamma} \\
m_\delta^4&\equiv&\frac{\pi^2 m_D^4 \xi^2 p_z^2 p_x^2 {\bf{p}}^2}{16({\bf{p}}^2 + \xi p_x ^2)^{3}}\, . \label{eq:mdelta}
\ea
\checked{mn}
The above expressions apply when ${\bf n}=(0,0,1)$ points along the $z$-axis and ${\bf p}$ lies in the $x-z$ plane; in the general case, $p_z$ and $p_x$ should be replaced by $\bf{p\cdot n}$ and $|\bf{p- (p\cdot n)n}|$, respectively.  With this, we can write down an expression for the real part of the potential which is valid to all orders in $\xi$ \cite{Dumitru:2007hy,Strickland:2011aa}
\ba
V({\bf{r}},\xi) &=& -g^2 C_F\int \frac{d^3{\bf{p}}}{(2\pi)^3} \,
\left( e^{i{\bf{p \cdot r}}} -1 \right) \tilde{\cal D}_R^{00}(\omega=0, \bf{p},\xi) \nonumber\\
&=& -g^2 C_F\int \frac{d^3{\bf{p}}}{(2\pi)^3} \,
\left( e^{i{\bf{p \cdot r}}} -1 \right) \frac{{\bf{p}}^2+m_\alpha^2+m_\gamma^2}
 {({\bf{p}}^2 + m_\alpha^2 +
     m_\gamma^2)({\bf{p}}^2+m_\beta^2)-m_\delta^4} \, ,
     \label{eq:repot}
\ea
\checked{mn}
where we have used the fact that, in this frame, $A^{00} = C^{00} = D^{00}=0$ and $B^{00} = {\bf p}^2/\omega^2$.  Unfortunately, from this point forward one must compute this integral numerically except in some special limiting cases \cite{Dumitru:2007hy,Strickland:2011aa}.

\section{The Feynman gluon self-energy}
Starting from the relation (\ref{eq:PiF0}) and using $S_{R,F,A}(k)=\slashed{k}\Delta_{R,A,F}(k)$  we have
\ba
\Pi^{\mu\nu}_F(p,\xi)=-2i e^2\int &&\!\!\!\frac{d^4k}{(2\pi)^4} \Big(q^\mu k^\nu+q^\nu k^\mu-\eta^{\mu\nu}q\cdot k\Big)\times\nonumber \\ &&\Big[\Delta_F(k)\Delta_F(q)
-\Big(\Delta_R(k)-\Delta_A(k)\Big)
\Big(\Delta_R(q)-
\Delta_A(q)\Big)\Big] \,.
\label{eq:PiF2}
\ea
\checked{m}
Using (\ref{eq:propag}) and
\be
\Delta_R(q)-\Delta_A(q)=-2\pi i\, \mbox{sgn}(q_0)\delta(q^2) \,,
\ee
\checked{m}
one obtains the term inside [...] in (\ref{eq:PiF2}) as
\be
[...]= (-2\pi i)^2 \delta(k^2)\delta(q^2)\Big[(1-2f_F({\bf k}))(1-2f_F({\bf q}))-{\rm sgn}(k_0){\rm sgn}(q_0)\Big]\,.
\ee
\checked{m}
Also by definition
\ba
\delta(k^2)=\frac{1}{2|{\bf k}|} \Big(\delta(k_0-|{\bf k}|)+\delta(k_0+|{\bf k}|)\Big)\,, \\
\delta(q^2)=\delta(k_0^2-{\bf k}^2+p^2-2k_0\omega+2{\bf k}\cdot{\bf p})\,.
\ea
\checked{m}
Using the relations above and performing the integral over $k_0$, and setting ${\bf k}\rightarrow -{\bf k}$ in the negative-energy contribution one has
\ba
\frac{\Pi^{\mu\nu}_F(p,\xi)}{2i\pi e^2}&=&  \int \frac{d^3{\bf k}}{(2\pi)^3|{\bf k}|} (q^\mu k^\nu+q^\nu k^\mu-\eta^{\mu\nu}q\cdot k)\delta(p^2-2|{\bf k}| \omega+2{\bf k\cdot p})\times \nonumber \\ &&\hspace{3cm}\Big[(1-2f_F({\bf k}))(1-2f_F({\bf q}))-{\rm sgn}(|{\bf k}|){\rm sgn}(|{\bf k}|-\omega) \Big] \Bigg|_{k_0 = |{\bf k}|}\nonumber \\
&+& \int \frac{d^3{\bf k}}{(2\pi)^3|{\bf k}|} (\bar{q}^\mu k^\nu+\bar{q}^\nu k^\mu-\eta^{\mu\nu}\bar{q}\cdot k)\delta(-p^2-2|{\bf k}| \omega+2{\bf k\cdot p})\times \nonumber \\ &&\hspace{3cm}\Big[(1-2f_F({\bf k}))(1-2f_F({\bf q}))-{\rm sgn}(|{\bf k}|){\rm sgn}(|{\bf k}|+\omega) \Big] \Bigg|_{k_0 = |{\bf k}|} , \;\;
\ea
\checked{m}
where   $\bar{q}\equiv k+p$.
Taking the HL limit one obtains the Feynman photon self-energy in the HL limit
\be 
\Pi^{\mu\nu}_F(p,\xi)= \frac{16i\pi e^2}{{\bf |p|}}\int \frac{d^3{\bf k}}{(2\pi)^3}
v^\mu v^\nu f_F({\bf k})(f_F({\bf k})-1)  \delta\left({\bf v\cdot {\hat p}}-\frac{\omega}{|{\bf p}|}\right)\,.
\label{eq:PiF3}
\ee
\checked{m}

By specializing to the anisotropic distribution function, we can simplify $\Pi_F^{\mu\nu}$ a bit more. Using an anisotropic Fermi-Dirac distribution (\ref{eq:FB-dist}) and Eq.~(\ref{eq:dist-identity2}) 
\ba
e^2\int_0^\infty d|{\bf k}|{\bf k}^2 f_F({\bf k})(f_F({\bf k})-1)&=&
\frac{\lambda e^2}{\sqrt{1+\xi ({\bf v\cdot n})^2}}\int_0^\infty d|{\bf k}|{\bf k}^2\frac{\partial f_F({\bf k})}{\partial|{\bf k}|}= \frac{-\lambda \pi^2m_D^2}{2(1+\xi ({\bf v\cdot n})^2)^{3/2}} \, , \;\;\;\;
\ea
\checked{m}
Using this identity, (\ref{eq:PiF3}) becomes
\be
\Pi^{\mu\nu}_F(p,\xi)=-\frac{i\lambda  m_D^2}{|{\bf p}|} \int d\Omega \frac{v^\mu v^\nu}{(1+\xi ({\bf v\cdot n})^2)^{3/2}} \delta\Big({\bf v\cdot \hat{p}}-\frac{\omega}{|\bf p|}\Big) \,.
\label{eq:PiF4}
\ee
\checked{m}
%

\subsection*{Feynman gluon self-energy}

Once more, the expression for the HL Feynman photon self-energy (\ref{eq:PiF4}) can be used to obtain the HL Feynman gluon self-energy provided that the Debye mass is replaced by (\ref{eq:qcd-debye}). The spatial block of Feynman gluon self-energy is
\be
\Pi^{ij}_F(p,\xi)=-\frac{i\lambda  m_D^2}{|{\bf p}|} \int d\Omega \frac{v^i v^j}{(1+\xi ({\bf v\cdot n})^2)^{3/2}} \delta\Big({\bf v\cdot \hat{p}}-\frac{\omega}{|\bf p|}\Big) \,.
\label{eq:PiFij}
\ee
\checked{m}
The integral above can be solved analytically as a function of ${\bf p}$, $\omega$, and $\xi$. For this purpose, it is more convenient to take ${\bf \hat{p}}$ along $z$-axis and ${\bf n}$ in the $x-z$ plane characterized by \mbox{$(\theta_n,\phi_n)=(\arctan(p_x/p_z),\pi)$}, where $p_x \equiv |{\bf p} - ({\bf p}\cdot{\bf n}) {\bf n}|$ and $p_z \equiv {\bf p} \cdot {\bf n}$ are the perpendicular and parallel components of ${\bf p}$ respect to ${\bf n}$ . Using this setup one has
\ba
{\bf n}&=&\left(-\frac{p_x}{|{\bf p}|},0,\frac{p_z}{|{\bf p}|}\right);\:\quad {\bf v}=(\sin\theta\cos\phi,\sin\theta\sin\phi,\cos\theta);\:\quad \hat{\bf p}=(0,0,1)\,,
\label{eq:def2}
\ea
\checked{m}
which gives
\ba
 {\bf v\cdot n}&=&-\frac{p_x}{|{\bf p}|}\sin\theta\cos\phi+\frac{p_z}{|{\bf p}|}\cos\theta\,, \\
 {\bf v\cdot \hat{p}}&=&\cos\theta\,.
\ea
Now by defining 
\be
{\bf u}=\left(\sqrt{1-\frac{\omega^2}{{\bf p}^2}}\cos\phi,\sqrt{1-\frac{\omega^2}{{\bf p}^2}}\sin\phi,\frac{\omega}{|{\bf p}|}\right)\,,
\ee
\checked{m}
and taking the integral over $\theta$ in (\ref{eq:PiFij}) we have
\be
\Pi^{ij}_F(p,\xi)=-\frac{i\lambda  m_D^2}{|{\bf p}|} \Theta(|{\bf p}|^2-\omega^2) \int d\phi \frac{{\bf u}_i {\bf u}_ j}{[1+\xi ({\bf u\cdot n})^2]^{3/2}} \,.
\ee
\checked{m}
Note that, for easier evaluation, the Feynman self-energy is calculated in a coordinate system that is different than the one used for retarded and advanced self-energies. Provided that we stay in the same frame until we obtain the scalars (which are coordinate independent), i.e. expansion coefficients of $\Pi^{\mu\nu}_F$, we do not need to do any further transformation to translate between the two choices of coordinate system. Therefore, for example, when needed for calculating $\Pi_F$ coefficients, i.e. Eqs.~(\ref{eq:abgd}), the tensor $A$ should be defined in the frame defined by (\ref{eq:def2}). Once we have obtained the coefficients through various tensor contractions, we can use the matrices  $A$, $B$, $C$, and $D$ in original frame to construct $\Pi_F^{\mu\nu}$.

The resulting HL tensor expansion coefficients for Feynman self-energy in the static limit are
\ba
\lim_{\omega \rightarrow 0} \alpha_F&=& -\frac{4 i \lambda m_D^2}{|{\bf p}| \varsigma} \Big[ E(-\varsigma) - K(-\varsigma) \Big] ,\nonumber \\
\lim_{\omega \rightarrow 0} \beta_F&=& -\frac{4 i \lambda m_D^2 \omega^2}{{\bf p}^3 (1+\varsigma)} E(-\varsigma) \, ,\nonumber \\
\lim_{\omega \rightarrow 0} \gamma_F&=& -\frac{4 i \lambda m_D^2}{|{\bf p}|} \bigg[ \frac{2}{\varsigma} K(-\varsigma) - \frac{2+\varsigma}{\varsigma(1+\varsigma)} E(-\varsigma) \bigg] ,\nonumber \\
\lim_{\omega \rightarrow 0} \delta_F&=& -\frac{4 i \lambda m_D^2 \omega^2 p_z }{{\bf p}^3 p_x^2 (1+\varsigma)} \bigg[ \frac{1-\varsigma}{1+\varsigma} E(-\varsigma) - K(-\varsigma) \bigg] ,
\ea
\checked{mny}
where $\varsigma \equiv \xi p_x^2/{\bf p}^2$, and $K$ and $E$ are complete elliptic integrals of the first and second kind, respectively, defined by 
\ba
K(x)&\equiv& \int _0^{\pi/2}\frac{1}{\sqrt{1-x\sin^2\phi}} \,d\phi\,,\nonumber \\
E(x)&\equiv& \int _0^{\pi/2}\sqrt{1-x\sin^2\phi}\,\,d\phi\,.
\ea
\checked{m}
In the isotropic case, the relations above simplify to
\ba
\lim_{\xi\rightarrow 0}  \lim_{\omega\rightarrow 0} \alpha_F &\rightarrow & -\frac{ i \pi \lambda m_D^2}{|{\bf p}|} \,, \nonumber \\
\lim_{\xi\rightarrow 0}  \lim_{\omega\rightarrow 0} \beta_F &\rightarrow & -\frac{2 i \pi \lambda m_D^2 \omega^2}{{\bf p}^3}\,, \nonumber \\
\lim_{\xi\rightarrow 0}  \lim_{\omega\rightarrow 0} \gamma_F &\rightarrow & 0\,, \nonumber \\
\lim_{\xi\rightarrow 0}  \lim_{\omega\rightarrow 0} \delta_F &\rightarrow & 0\,.
\ea
\checked{mn}
The first two agree with the isotropic results given by Eqs.~(20) and (19) of Ref.~\cite{Carrington:1997sq} for $\Pi^T_F$ and $\Pi^L_F$, respectively, upon using $\beta_F = (\omega^2/{\bf p}^2) \Pi^L_F$ \cite{Romatschke:2003ms} and $m^2_D$ is replaced by its QED definition, i.e. Eq.~(\ref{eq:mDQED}).

\section{The hard-loop Feynman propagator}

In this section, we obtain the static limit ($\omega\rightarrow 0$) of $\tilde{\cal D}^{00}_F$ using Eq.~(\ref{eq:2b8})
\ba
\tilde{\cal D}_{F}(p)= && (1+2f_B({\bf p}))\, \mbox{sgn}(\omega)\,
[\tilde{\cal D}_{R}(p)-\tilde{\cal D}_{A}(p)]
\nonumber \\
&& +\tilde{\cal D}_{R}(p)\,\{\Pi _F(p)-(1+2f_B({\bf p}))\, \mbox{sgn}(\omega)\, [\Pi
_R(p)-\Pi _A(p)]\} \,  \tilde{\cal D}_A(p)~. \label{eq:2b88}
\ea
\checked{m}
Taking the $00$ component using (\ref{eq:ABCD}) and (\ref{eq:Dexp}), the first term becomes
\be
 \left[1+2f_B({\bf p})\right]\, \mbox{sgn}(\omega) (\tilde{\cal D}_R^{00}-\tilde{\cal D}_A^{00})=\left[1+2f_B({\bf p})\right]\, \mbox{sgn}(\omega) (\beta'_R-\beta'_A)\frac{{\bf p}^2}{\omega^2}\,,
 \label{eq:trm1}
\ee
\checked{m}
where considering (\ref{eq:ABCD}) we have used $A^{00}=C^{00}=D^{00}=0$ and $B^{00} = {\bf p}^2/\omega^2$.

Using (\ref{eq:PiRcoef}), one can write the tensor basis coefficients of $\Pi_{R/A}$ as
\ba
\alpha_{R/A}&=&\alpha_0 \pm  i \omega \alpha_1\,, \nonumber\\
\beta_{R/A}&=&(\beta_0 \pm  i \omega \beta_1)\omega^2\,,\nonumber \\
\gamma_{R/A}&=&\gamma_0 \pm  i \omega \gamma_1\,, \nonumber \\
\delta_{R/A}&=&(\pm i\delta_0 + \omega \delta_1)\omega\,,
\label{eq:PiRcoef2}
\ea
\checked{m}
where all new coefficients with subscripts `0' and `1' are independent of $\omega$ and can be easily read off from (\ref{eq:PiRcoef}). Also, specializing to the anisotropic distribution function, Eq.~(\ref{eq:FB-dist}), one has
\be
\lim_{\omega \rightarrow 0} (1+2f_B({\bf p}))\, \mbox{sgn}(\omega) \approx \frac{2 \lambda}{\sqrt{1+\xi {\bf (v\cdot n)}^2}}\frac{1}{\omega}+{\cal O}(\omega^0) \, .
\ee
\checked{m}
Using relations (\ref{eq:coefPrime}) for $\beta'_{R/A}$, one finds  $ (1+2f_B)\, \mbox{sgn}(\omega)(\tilde{\cal D}_R^{00}-\tilde{\cal D}_A^{00})$ at leading order in the static limit 
\ba
 &&\lim_{\omega \rightarrow 0} (1+2f_B({\bf p}))\mbox{sgn}(\omega)(\tilde{\cal D}_R^{00}-\tilde{\cal D}_A^{00})= \nonumber \\
 &&\hspace{10mm}\frac{4i\lambda}{{\bf p}^2\sqrt{1+\xi {\bf (v\cdot n)}^2}}\frac{-p_x^2(\alpha_1+\gamma_1)\delta_0^2+({\bf p}^2+\alpha_0+\gamma_0)
 \left(\beta_1({\bf p}^2+\alpha_0+\gamma_0)-2p_x^2\delta_0\delta_1\right)}{\left((\beta_0-1)({\bf p}^2+\alpha_0+\gamma_0)+p_x^2\delta_0^2\right)^2}\;\,.
  \;\;\;\; \label{eq:term1}
\ea
\checked{mn}
Now, we turn to the next term, $(\tilde{\cal D}_R\Pi_F \tilde{\cal D}_A)^{00}$ which, in the static limit, becomes
\ba
&&\lim_{\omega\rightarrow 0}(\tilde{\cal D}_R \Pi_F \tilde{\cal D}_A)^{00}= \nonumber\\&&
\hspace{1cm}\frac{-4i \lambda m_D^2}{[(\beta_0-1)({\bf p}^2+\alpha_0+\gamma_0)+p_x^2 \delta_0^2]^2}\!\!
\left[ \frac{ \xi ( {\bf p}^2+\alpha_0+\gamma_0)^2 - \delta_0^2 {\bf p}^4 }{\xi {\bf p}^3 ({\bf p}^2 + \xi p_x^2)}  E\!\left(-\frac{\xi p_x^2}{{\bf p}^2} \right) +
\frac{\delta_0^2}{|{\bf p}|\xi} K\!\left(-\frac{\xi p_x^2}{{\bf p}^2} \right) 
\right] \!.  \hspace{1cm} \label{eq:term2}
\ea
\checked{mn}
The last term can be calculated similarly using the setup above and relations listed in App.~\ref{app:tensor-relations}
\ba
&&\lim_{\omega\rightarrow 0}(1+2f_B({\bf p}))\mbox{sgn}(\omega)[\tilde{\cal D}_R(\Pi_R-\Pi_A)\tilde{\cal D}_A]^{00}=\nonumber \\
&&\hspace{10mm}\frac{-4i\lambda}{{\bf p}^2\sqrt{1+\xi {\bf (v\cdot n)}^2}}\frac{-p_x^2(\alpha_1+\gamma_1)\delta_0^2+({\bf p}^2+\alpha_0+\gamma_0)
 \Big(\beta_1({\bf p}^2+\alpha_0+\gamma_0)-2p_x^2\delta_0\delta_1\Big)}{\Big((\beta_0-1)({\bf p}^2+\alpha_0+\gamma_0)+p_x^2\delta_0^2\Big)^2}\,.
\;\;\;\;\;  \label{eq:term3}
\ea
\checked{mn}
From the relations above (\ref{eq:term1}) and (\ref{eq:term3}), one can see that the first and the last terms of Eq.~(\ref{eq:2b88}) cancel each other leaving the second term which, using the parametrizations (\ref{eq:malpha})-(\ref{eq:mdelta}),  gives
\ba
&&\lim_{\omega\rightarrow 0}\tilde{\cal D}^{00}_{F}(p,\xi)= \nonumber\\
&&\hspace{7mm} \frac{4i \lambda m_D^2}{\varsigma{|\bf p}|[({\bf p}^2+m_\beta^2)({\bf p}^2 + m_\alpha^2 + m_\gamma^2)-m_\delta^4]^2}
\left[ \frac{m_\delta^4- \varsigma ( {\bf p}^2+m_\alpha^2 + m_\gamma^2)^2 }{1+\varsigma}  E\!\left(-\varsigma \right) -
m_\delta^4 K\!\left(-\varsigma \right) 
\right]\!. \hspace{8mm} \label{eq:term2alt}
\ea
\checked{mny}
This is our final result for the static Feynman propagator.  Expanding our final result in terms of powers of $\xi$
\be
\lim_{\omega\rightarrow 0} \tilde{\cal D}^{00}_{F}(p,\xi) = - \frac{2 \pi i m_D^2 \lambda}{|{\bf p}| ({\bf p}^2 + m_D^2)^2}+\frac{i\pi m_D^2 \lambda \xi}{6 {\bf p}^3({\bf p}^2+m_D^2)^3}\Big[9{\bf p}^2p_x^2+m_D^2(8{\bf p}^2-15p_x^2)\Big]+{\cal O}(\xi^2) \, ,
\ee
\checked{n}
which is in agreement with earlier results obtained in the small-$\xi$ limit \cite{Laine:2006ns,Burnier:2009yu,Dumitru:2009fy}.  Note that, if one expands (\ref{eq:term2alt}) to higher order in $\xi$, one finds increasingly negative powers of $|{\bf p}|$ which result in infrared divergences in the corresponding corrections to the imaginary part of the static heavy quark potential.  The full result (\ref{eq:term2alt}) is, however, infrared safe. 

\subsection*{Pinch singularity}

As mentioned in the introduction, the imaginary part of the heavy-quark potential can be obtained from the Fourier transform of the static limit of the Feynman propagator
\be
V_I({\bf r},\xi) \equiv - \frac{g^2 C_F}{2} \int \frac{d^3{\bf p}}{(2\pi)^3} \left(e^{i {\bf p}\cdot {\bf r}} - 1 \right) 
\tilde{\cal D}^{00}_F \Big|_{\omega \rightarrow 0} \; .
\ee
However, (\ref{eq:term2alt}) contains a pinch singularity which is related to the presence of the (chromo-)Weibel instability in momentum-space anisotropic plasmas \cite{Mrowczynski:2000ed,Romatschke:2003ms}.  This pinch singularity causes the imaginary-part of the potential to be ill-defined.  To see that this is the case, we point out that (\ref{eq:term2alt}) can be written more compactly using
\be
({\bf p}^2+m_\beta^2)({\bf p}^2 + m_\alpha^2 + m_\gamma^2)-m_\delta^4 = ({\bf  p}^2 + m_+^2)({\bf  p}^2 + m_-^2) \, ,
\ee
\checked{m}
where
\be
2 m_{\pm}^2 = M^2 \pm \sqrt{M^4-4[m_\beta^2(m_\alpha^2+m_\gamma^2)-m_\delta^4]} \; ,
\label{mpm}
\ee
\checked{m}
with $M^2 = m_\alpha^2+m_\beta^2+m_\gamma^2$ \cite{Romatschke:2003ms}.  One can show that $m_+^2$ is positive for all $\xi$ and angles of propagation; however, $m_-^2$ can be negative for some propagation angles.  This is illustrated in Fig.~2 of Ref.~\cite{Romatschke:2003ms} and discussed in the surrounding text.  As a result, in unstable regions of phase space, ${\bf p}^2 + m_-^2$ can go to zero.  This occurs already in the integral necessary to obtain the real part of the potential (\ref{eq:repot}); however, in this case there is only one power of ${\bf p}^2 + m_-^2$ in the denominator, which results in a simple pole that can be integrated using a principle part prescription, e.g. ${\bf p}^2 + m_-^2 \rightarrow (|{\bf p}| + |m_-|)(|{\bf p}| - |m_-|)$.  In the case of $V_I$, however, the denominator of the integrand contains $({\bf  p}^2 + m_-^2)^2$, which results in a double pole in the Fourier transform.   To see that this is, in fact, a pinch singularity we note that the prefactor of (\ref{eq:term2alt}) which causes the trouble comes from the product of retarded and advanced propagators, $\Delta_{G}^R \Delta_{G}^A$.  Keeping track of the $i \epsilon$'s, one finds two simple poles shifted by $\pm i\epsilon$ which collapse onto the real axis as $\epsilon \rightarrow 0$, forming a double pole.

\section{Conclusions and outlook}

In this paper we presented a calculation of the hard-loop resummed retarded, advanced, and Feynman (symmetric) gluon propagators in a momentum-space anisotropic plasma with a single anisotropy direction, ${\bf n}$.  We used the real-time formalism throughout and, when available, we compared to previously obtained results.  Our main new result is an expression for the Feynman gluon propagator which is accurate to all orders in the anisotropy parameter $\xi$ (\ref{eq:term2alt}).  Unlike results obtained using Taylor expansion in $\xi$, (\ref{eq:term2alt}) is infrared finite, however, it possesses a pinch singularity which formally renders the imaginary part of the heavy-quark potential infinite.   The existence of this pinch singularity can be traced back to the existence of unstable modes in a momentum-space anisotropic quark-gluon plasma \cite{Mrowczynski:2000ed,Romatschke:2003ms,Mrowczynski:2016etf}.  

A pinch-singularity emerges because one presumes that the collective modes, which are determined through a linearized treatment, apply at all times.  In an equilibrium (stable) situation the field amplitudes are bounded (and small) and such a treatment makes some sense.  However, in our case, the system is unstable and some subset of the linearized collective modes grow exponentially for all times, which upon taking the static limit ($\omega \rightarrow 0$ or  $t \rightarrow \infty$) results in an infinite effect.  As a result, in the presence of unstable modes this scheme is ill-defined and it seems necessary to impose an upper time limit for unstable mode growth.  At the most conservative, the upper time limit for unstable mode growth would be set by the lifetime of the QGP, however, in practice one finds that plasma instabilities may saturate on a shorter timescale.  

In terms of the calculation presented herein, one could attempt to implement the physics of instability saturation or finite plasma lifetime by imposing an infrared cutoff on the frequency $\omega_0 \sim {\rm max}(\tau^{-1}_{\rm instability},\tau^{-1}_{\rm QGP})$ where $\tau_{\rm instability}$ is the expected timescale for the saturation of unstable field growth and $\tau_{\rm QGP} \sim $ 10 fm/c $\sim$ \mbox{1/(20 MeV)} is the typical lifetime of the quark-gluon plasma.
Detailed simulations of anisotropic non-abelian plasmas in fixed boxes show that unstable exponential growth terminates when the gauge field amplitude reaches the soft scale and the subsequent gauge field dynamics transform into a much slower turbulent cascade of energy from soft scales to hard scales \cite{Arnold:2005ef,Arnold:2005qs,Rebhan:2005re,Strickland:2007fm,Dumitru:2006pz}.  More recent studies of chromo-Weibel dynamics in an expanding non-Abelian plasmas found that unstable modes saturate on a time scale of 3-4 fm/c at LHC energies \cite{Rebhan:2008uj,Attems:2012js}.  Combined with the QGP lifetime estimate, one has $\omega_0 \sim 20 - 70 $ MeV.  In practice, an infrared cutoff such as this will lift the poles off the real axis, even in the limit $\epsilon \rightarrow 0$ due to the finite imaginary linear correction in $\omega$ to the structure functions $\alpha_{R/A}$, $\beta_{R/A}$, $\gamma_{R/A}$, and $\delta_{R/A}$ listed in (\ref{eq:PiRcoef}).  

While such a phenomenological prescription may work in practice, it introduces a fundamental problem, since the 00-component gluon propagator which is used to define the potential is not gauge invariant for finite $\omega$ [see Eq.~(\ref{eq:tildeDRA})].  For ``reasonable gauges'' the dependence may not be large, but nevertheless this is an unsatisfactory resolution of this problem on general grounds.  For this reason, one should simultaneously pursue the possibility to measure the potential numerically using classical gauge theory simulations similar to those used to measure the imaginary part of the heavy quark potential in the equilibrium limit \cite{Laine:2007qy}.  In this method, one determines the imaginary part of the potential by measuring the classical Wilson loop which amounts to a two-point correlation function of two spatial Wilson lines.  With this method one would be able to obtain a gauge-invariant imaginary part of the potential, however, one still would not be able to take the $t \rightarrow \infty$ limit due to finite computational resources, break-down of the classical hard loop limit, etc.  

Finally, as another path forward, one might consider adding the effect of collisions in the computation of the anisotropic structure functions.  Previous studies \cite{Schenke:2006xu} have shown that at fixed $\xi$, if the collision rate exceeds a certain threshold, then unstable modes are eliminated from the spectrum.  This would provide another way to regulate/eliminate the ill-defined effect of unstable modes in the heavy-quark potential.  We leave the investigation of these possibilities for future work(s).

\acknowledgments
M. Nopoush and M. Strickland were supported by the U.S. Department of Energy, Office of Science, Office of Nuclear Physics under Award No.~DE-SC0013470.  Y. Guo gratefully acknowledges support from the NSFC of China under Project Nos.~11665008 and 11647309, the Natural Science Foundation of Guangxi Province of China under Project No.~2016GXNSFFA380014, and the ``Hundred Talents Plan'' of Guangxi Province of China.

\appendix

\section{Tensor relations}
\label{app:tensor-relations}
In this appendix, we present various tensor identities obeyed by our basis tensors (\ref{eq:ABCD}). Useful identities for the contraction of any two basis tensors are as following
\ba
A\cdot A &=& -A\,, \\
B\cdot B &=& -Z B\,, \\
C\cdot C&=& -C\,,\\
D\cdot D &=&-p_\perp^2 (B+Z C)\,,\\
A\cdot B&=& B\cdot A = 0\,,\\
A\cdot C&=& C \cdot A = -C\,,\\
B\cdot C&=& C \cdot B = 0\,,\\
A\cdot D&=&C\cdot D, \\
D\cdot A&=& D\cdot C\, , \\
D\cdot B&=&Z A\cdot D\, ,\\
B\cdot D&=&Z D\cdot C\,,\\
A\cdot D+D\cdot A &=&-D\,,\\
B\cdot D+D\cdot B&=&-Z D\,,\\
C\cdot D+D\cdot C&=&-D\,,
\ea
\checked{m}
with $Z\equiv p^2/\omega^2=1- {\bf p}^2/\omega^2$. Out of 64 possible contractions of any three tensors of $A$, $B$, $C$, and $D$ the non-trivial ones are
\ba 
A\cdot A \cdot A=A\,,  \\
\frac{B\cdot B\cdot B}{Z^2}=\frac{B\cdot D\cdot D}{Zp_\perp^2}=\frac{D\cdot D\cdot B}{Z p_\perp^2}=\frac{D\cdot A\cdot D}{p_\perp^2}=\frac{D\cdot C\cdot D}{p_\perp^2}= B\,, \\
A\cdot A\cdot C=A\cdot C\cdot C=A\cdot C\cdot A=C\cdot A\cdot A=C\cdot C\cdot C=C\cdot A\cdot C=C\cdot C\cdot A=C\,,  \\
A\cdot D\cdot D=C\cdot D\cdot D=D\cdot D\cdot A=D\cdot D\cdot C=\frac{D\cdot B\cdot D}{Z}= Zp_\perp^2 C\,,  \\
D\cdot D\cdot D=Zp_\perp^2D\,, \\
A\cdot A\cdot D=C\cdot C\cdot D=C\cdot A\cdot D=A\cdot C\cdot D=\frac{D\cdot B\cdot B}{Z^2}=\frac{A\cdot D\cdot B}{Z}=\frac{C\cdot D\cdot B}{Z}\,, \label{eq:AAD}\\
D\cdot A\cdot A=D\cdot C\cdot C=D\cdot A\cdot C=D\cdot C\cdot A=\frac{B\cdot B\cdot D}{Z^2}=\frac{B\cdot D\cdot A}{Z}=\frac{B\cdot D\cdot C}{Z}\,, \label{eq:DAA} \\
B \cdot  D \cdot B = A \cdot D \cdot A = C \cdot D \cdot C = A \cdot D \cdot C = C \cdot D \cdot A = 0 \, ,
\ea
\checked{m}
where (\ref{eq:AAD}) and (\ref{eq:DAA}) contain the contractions that cannot be expressed in terms of any single basis tensors. In all relations above $p_\perp\equiv |{\bf p}-({\bf p}\cdot {\bf n}){\bf n}|$ is the component of ${\bf p}$ perpendicular to ${\bf n}$.
Using the identities listed above, one can calculate the contraction of any two dressed propagators with a self-energy in between, $\tilde{\cal D}_1\Pi_2 \tilde{\cal D}_3$, as 
\ba
\tilde{\cal D}_1\Pi_2 \tilde{\cal D}_3&=&\alpha'_1\alpha_2\alpha'_3 A \nonumber \\
&+&\Big[\beta'_1\beta_2\beta'_3 Z^2+\beta'_1\delta_2\delta'_3p_\perp^2Z+\delta'_1\delta_2\beta'_3p_\perp^2Z+
\delta'_1\alpha_2\delta'_3p_\perp^2+ \delta'_1\gamma_2\delta'_3p_\perp^2\Big]B\nonumber\\
&+&\Big[\delta'_1\beta_2\delta'_3p_\perp^2Z^2+\delta'_1\delta_2\gamma'_3 p_\perp^2Z+\gamma'_1\delta_2\delta'_3 p_\perp^2Z+\alpha'_1\delta_2\delta'_3p_\perp^2Z
+\delta'_1\delta_2\alpha'_3p_\perp^2Z
+\alpha'_1\alpha_2\gamma'_3\nonumber \\&+&\gamma'_1\alpha_2\alpha'_3
+\alpha'_1\gamma_2\alpha'_3+\gamma'_1
(\gamma_2+\alpha_2)\gamma'_3+\alpha'_1\gamma_2\gamma'_3+\gamma'_1\gamma_2\alpha'_3\Big]C +\delta'_1\delta_2\delta'_3Z p_\perp^2 D\nonumber \\
&-& \Big[\beta'_1\beta_2\delta'_3 Z^2+\beta'_1\delta_2\alpha'_3 Z+\beta'_1\delta_2\gamma'_3 Z+\delta'_1\alpha_2\alpha'_3+
\delta'_1\alpha_2\gamma'_3 +\delta'_1\gamma_2\alpha'_3+\delta'_1\gamma_2\gamma'_3\Big]D\cdot A \nonumber \\
&-&\Big[\delta'_1\beta_2\beta'_3Z^2+\alpha'_1\delta_2\beta'_3Z
+\gamma'_1\delta_2\beta'_3Z+\alpha'_1\alpha_2\delta'_3+\gamma'_1\gamma_2\delta'_3
+\alpha'_1\gamma_2\delta'_3
+\gamma'_1\alpha_2\delta'_3\Big]A\cdot D \, .\hspace{1cm} \ea
\checked{m}

\section{An alternative method for calculating the expansion coefficients}
\label{app:alternative}

In this section, we present an alternative method to Eq.~(\ref{eq:abgd}) for calculating the expansion coefficients. This method, in comparison to the previous one, is based on the four-vector form of tensors $A$, $B$, $C$, and $D$. Any arbitrary rank-2 tensor ${\cal T}$ can be expanded in terms of basis tensors as
\ba
{\cal T}=\alpha A+\beta B+\gamma C+\delta D\,. 
\ea
\checked{m}
Let's start with calculating the following terms using the identities presented in App.~\ref{app:tensor-relations} 
\ba
\frac{1}{2}(A\cdot {\cal T}+{\cal T}\cdot A)&=&
-\alpha A-\gamma C-\frac{1}{2}\delta D\,, \\
\frac{1}{2}(B\cdot {\cal T}+{\cal T}\cdot B)&=&-Z \beta B-\frac{1}{2}Z\delta D\,, 
\\
\frac{1}{2}(C\cdot {\cal T}+{\cal T}\cdot C)&=&-(\alpha+\gamma) C-\frac{1}{2}\delta D\,, 
\\
\frac{1}{2}(D\cdot {\cal T}+{\cal T}\cdot D)&=&-\frac{1}{2}(\alpha+Z\beta +\gamma)D-Zp_\perp^2\delta C-p_\perp^2\delta B \,.
\ea
\checked{m}
Using 
\ba
{\rm Tr} [A]&=&-2\,, \\
{\rm Tr} [B]&=&-Z\,, \\
{\rm Tr} [C]&=&-1\,, \\
{\rm Tr} [D]&=&0\,, 
\ea
\checked{m}
we have 
\ba
{\rm Tr}\Big[{\textstyle \frac{1}{2}}(A\cdot {\cal T}+{\cal T}\cdot A)\Big]&=& {\rm Tr}(A\cdot {\cal T})=
2\alpha +\gamma\,, \\
{\rm Tr}\Big[{\textstyle \frac{1}{2}}(B\cdot {\cal T}+{\cal T}\cdot B)\Big]&=& {\rm Tr}(B\cdot {\cal T})=
Z^2\beta\,, \\
{\rm Tr}\Big[{\textstyle \frac{1}{2}}(C\cdot {\cal T}+{\cal T}\cdot C)\Big]&=& {\rm Tr}(C\cdot {\cal T})=
\alpha +\gamma\,,\\
{\rm Tr}\Big[{\textstyle \frac{1}{2}}(D\cdot {\cal T}+{\cal T}\cdot D)\Big]&=& {\rm Tr}(D\cdot {\cal T})=
2Z p_\perp^2\delta\,,
\ea
\checked{m}
where we have used the fact that the trace of inner product of each the $A$, $B$, $C$, and $D$ tensors with an arbitrary tensor ${\cal T}$ is commutative. Solving for the expansion coefficients we have
\ba 
\alpha&=&{\rm Tr}(A\cdot {\cal T})-{\rm Tr}(C\cdot {\cal T})\,,\\
\beta&=&\frac{1}{Z^2}{\rm Tr}(B\cdot {\cal T})\,,\\
\gamma&=&2{\rm Tr}(C\cdot {\cal T})-{\rm Tr}(A\cdot {\cal T})\,,\\
\delta&=&\frac{1}{2Z p_\perp^2}{\rm Tr}(D\cdot {\cal T})\,.
\ea
\checked{mn}
As an example of usage of the above formul\ae, one can set ${\cal T} \equiv \Pi_{R,A,F}$ to obtain the expansion coefficients for the retarded, advanced, and Feynman self-energies directly from the their four-tensor expressions prior to restricting to the space-like components.

\bibliographystyle{JHEP}
\bibliography{gprop}

\providecommand{\href}[2]{#2}\begingroup\raggedright\begin{thebibliography}{10}

\bibitem{Mocsy:2013syh}
A.~Mocsy, P.~Petreczky and M.~Strickland, \emph{{Quarkonia in the Quark Gluon
  Plasma}}, \href{http://dx.doi.org/10.1142/S0217751X13400125}{\emph{Int. J.
  Mod. Phys.} {\bfseries A28} (2013) 1340012},
  [\href{https://arxiv.org/abs/1302.2180}{{\ttfamily 1302.2180}}].

\bibitem{Andronic:2015wma}
A.~Andronic et~al., \emph{{Heavy-flavour and quarkonium production in the LHC
  era: from proton--proton to heavy-ion collisions}},
  \href{http://dx.doi.org/10.1140/epjc/s10052-015-3819-5}{\emph{Eur. Phys. J.}
  {\bfseries C76} (2016) 107},
  [\href{https://arxiv.org/abs/1506.03981}{{\ttfamily 1506.03981}}].

\bibitem{Strickland:2014pga}
M.~Strickland, \emph{{Anisotropic Hydrodynamics: Three lectures}},
  \href{http://dx.doi.org/10.5506/APhysPolB.45.2355}{\emph{Acta Phys. Polon.}
  {\bfseries B45} (2014) 2355--2394},
  [\href{https://arxiv.org/abs/1410.5786}{{\ttfamily 1410.5786}}].

\bibitem{Florkowski:2010cf}
W.~Florkowski and R.~Ryblewski, \emph{{Highly-anisotropic and
  strongly-dissipative hydrodynamics for early stages of relativistic heavy-ion
  collisions}},
  \href{http://dx.doi.org/10.1103/PhysRevC.83.034907}{\emph{Phys.Rev.}
  {\bfseries C83} (2011) 034907},
  [\href{https://arxiv.org/abs/1007.0130}{{\ttfamily 1007.0130}}].

\bibitem{Martinez:2010sc}
M.~Martinez and M.~Strickland, \emph{{Dissipative Dynamics of Highly
  Anisotropic Systems}},
  \href{http://dx.doi.org/10.1016/j.nuclphysa.2010.08.011}{\emph{Nucl. Phys.}
  {\bfseries A848} (2010) 183--197},
  [\href{https://arxiv.org/abs/1007.0889}{{\ttfamily 1007.0889}}].

\bibitem{Martinez:2012tu}
M.~Martinez, R.~Ryblewski and M.~Strickland, \emph{{Boost-Invariant
  (2+1)-dimensional Anisotropic Hydrodynamics}}, {\emph{Phys.Rev.} {\bfseries
  C85} (2012) 064913}, [\href{https://arxiv.org/abs/1204.1473}{{\ttfamily
  1204.1473}}].

\bibitem{Ryblewski:2012rr}
R.~Ryblewski and W.~Florkowski, \emph{{Highly-anisotropic hydrodynamics in 3+1
  space-time dimensions}},
  \href{http://dx.doi.org/10.1103/PhysRevC.85.064901}{\emph{Phys. Rev.}
  {\bfseries C85} (2012) 064901},
  [\href{https://arxiv.org/abs/1204.2624}{{\ttfamily 1204.2624}}].

\bibitem{Bazow:2013ifa}
D.~Bazow, U.~W. Heinz and M.~Strickland, \emph{{Second-order (2+1)-dimensional
  anisotropic hydrodynamics}},
  \href{http://dx.doi.org/10.1103/PhysRevC.90.054910}{\emph{Phys.Rev.}
  {\bfseries C90} (2014) 054910},
  [\href{https://arxiv.org/abs/1311.6720}{{\ttfamily 1311.6720}}].

\bibitem{Tinti:2013vba}
L.~Tinti and W.~Florkowski, \emph{{Projection method and new formulation of
  leading-order anisotropic hydrodynamics}},
  \href{http://dx.doi.org/10.1103/PhysRevC.89.034907}{\emph{Phys.Rev.}
  {\bfseries C89} (2014) 034907},
  [\href{https://arxiv.org/abs/1312.6614}{{\ttfamily 1312.6614}}].

\bibitem{Nopoush:2014pfa}
M.~Nopoush, R.~Ryblewski and M.~Strickland, \emph{{Bulk viscous evolution
  within anisotropic hydrodynamics}},
  \href{http://dx.doi.org/10.1103/PhysRevC.90.014908}{\emph{Phys.Rev.}
  {\bfseries C90} (2014) 014908},
  [\href{https://arxiv.org/abs/1405.1355}{{\ttfamily 1405.1355}}].

\bibitem{Tinti:2015xwa}
L.~Tinti, \emph{{Anisotropic matching principle for the hydrodynamic
  expansion}}, \href{http://dx.doi.org/10.1103/PhysRevC.94.044902}{\emph{Phys.
  Rev.} {\bfseries C94} (2016) 044902},
  [\href{https://arxiv.org/abs/1506.07164}{{\ttfamily 1506.07164}}].

\bibitem{Bazow:2015cha}
D.~Bazow, U.~W. Heinz and M.~Martinez, \emph{{Nonconformal viscous anisotropic
  hydrodynamics}},
  \href{http://dx.doi.org/http://dx.doi.org/10.1103/PhysRevC.91.064903}{\emph{Phys.Rev.}
  {\bfseries C91} (2015) 064903},
  [\href{https://arxiv.org/abs/1503.07443}{{\ttfamily 1503.07443}}].

\bibitem{Strickland:2015utc}
M.~Strickland, M.~Nopoush and R.~Ryblewski, \emph{{Anisotropic hydrodynamics
  for conformal Gubser flow}},
  \href{http://dx.doi.org/10.1016/j.nuclphysa.2016.02.014}{\emph{Nucl. Phys.}
  {\bfseries A956} (2016) 268--271},
  [\href{https://arxiv.org/abs/1512.07334}{{\ttfamily 1512.07334}}].

\bibitem{Alqahtani:2015qja}
M.~Alqahtani, M.~Nopoush and M.~Strickland, \emph{{Quasiparticle equation of
  state for anisotropic hydrodynamics}},
  \href{http://dx.doi.org/10.1103/PhysRevC.92.054910}{\emph{Phys. Rev.}
  {\bfseries C92} (2015) 054910},
  [\href{https://arxiv.org/abs/1509.02913}{{\ttfamily 1509.02913}}].

\bibitem{Molnar:2016vvu}
E.~Molnar, H.~Niemi and D.~H. Rischke, \emph{{Derivation of anisotropic
  dissipative fluid dynamics from the Boltzmann equation}},
  \href{http://dx.doi.org/10.1103/PhysRevD.93.114025}{\emph{Phys. Rev.}
  {\bfseries D93} (2016) 114025},
  [\href{https://arxiv.org/abs/1602.00573}{{\ttfamily 1602.00573}}].

\bibitem{Molnar:2016gwq}
E.~Molnar, H.~Niemi and D.~H. Rischke, \emph{{Closing the equations of motion
  of anisotropic fluid dynamics by a judicious choice of a moment of the
  Boltzmann equation}},
  \href{http://dx.doi.org/10.1103/PhysRevD.94.125003}{\emph{Phys. Rev.}
  {\bfseries D94} (2016) 125003},
  [\href{https://arxiv.org/abs/1606.09019}{{\ttfamily 1606.09019}}].

\bibitem{Alqahtani:2016rth}
M.~Alqahtani, M.~Nopoush and M.~Strickland, \emph{{Quasiparticle anisotropic
  hydrodynamics for central collisions}},
  \href{http://dx.doi.org/10.1103/PhysRevC.95.034906}{\emph{Phys. Rev.}
  {\bfseries C95} (2017) 034906},
  [\href{https://arxiv.org/abs/1605.02101}{{\ttfamily 1605.02101}}].

\bibitem{Alqahtani:2017jwl}
M.~Alqahtani, M.~Nopoush, R.~Ryblewski and M.~Strickland, \emph{{3+1d
  quasiparticle anisotropic hydrodynamics for ultrarelativistic heavy-ion
  collisions}},  \href{https://arxiv.org/abs/1703.05808}{{\ttfamily
  1703.05808}}.

\bibitem{Romatschke:2003ms}
P.~Romatschke and M.~Strickland, \emph{{Collective modes of an anisotropic
  quark gluon plasma}},
  \href{http://dx.doi.org/10.1103/PhysRevD.68.036004}{\emph{Phys. Rev.}
  {\bfseries D68} (2003) 036004},
  [\href{https://arxiv.org/abs/hep-ph/0304092}{{\ttfamily hep-ph/0304092}}].

\bibitem{Romatschke:2004jh}
P.~Romatschke and M.~Strickland, \emph{{Collective modes of an anisotropic
  quark-gluon plasma II}},
  \href{http://dx.doi.org/10.1103/PhysRevD.70.116006}{\emph{Phys. Rev.}
  {\bfseries D70} (2004) 116006},
  [\href{https://arxiv.org/abs/hep-ph/0406188}{{\ttfamily hep-ph/0406188}}].

\bibitem{Kasmaei:2016apv}
B.~S. Kasmaei, M.~Nopoush and M.~Strickland, \emph{{Quark self-energy in an
  ellipsoidally anisotropic quark-gluon plasma}},
  \href{http://dx.doi.org/10.1103/PhysRevD.94.125001}{\emph{Phys. Rev.}
  {\bfseries D94} (2016) 125001},
  [\href{https://arxiv.org/abs/1608.06018}{{\ttfamily 1608.06018}}].

\bibitem{Dumitru:2007hy}
A.~Dumitru, Y.~Guo and M.~Strickland, \emph{{The Heavy-quark potential in an
  anisotropic (viscous) plasma}},
  \href{http://dx.doi.org/10.1016/j.physletb.2008.02.048}{\emph{Phys. Lett.}
  {\bfseries B662} (2008) 37--42},
  [\href{https://arxiv.org/abs/0711.4722}{{\ttfamily 0711.4722}}].

\bibitem{Dumitru:2009ni}
A.~Dumitru, Y.~Guo, A.~Mocsy and M.~Strickland, \emph{{Quarkonium states in an
  anisotropic QCD plasma}},
  \href{http://dx.doi.org/10.1103/PhysRevD.79.054019}{\emph{Phys. Rev.}
  {\bfseries D79} (2009) 054019},
  [\href{https://arxiv.org/abs/0901.1998}{{\ttfamily 0901.1998}}].

\bibitem{Burnier:2009yu}
Y.~Burnier, M.~Laine and M.~Vepsalainen, \emph{{Quarkonium dissociation in the
  presence of a small momentum space anisotropy}},
  \href{http://dx.doi.org/10.1016/j.physletb.2009.05.067}{\emph{Phys. Lett.}
  {\bfseries B678} (2009) 86--89},
  [\href{https://arxiv.org/abs/0903.3467}{{\ttfamily 0903.3467}}].

\bibitem{Dumitru:2009fy}
A.~Dumitru, Y.~Guo and M.~Strickland, \emph{{The Imaginary part of the static
  gluon propagator in an anisotropic (viscous) QCD plasma}},
  \href{http://dx.doi.org/10.1103/PhysRevD.79.114003}{\emph{Phys. Rev.}
  {\bfseries D79} (2009) 114003},
  [\href{https://arxiv.org/abs/0903.4703}{{\ttfamily 0903.4703}}].

\bibitem{Du:2016wdx}
Q.~Du, A.~Dumitru, Y.~Guo and M.~Strickland, \emph{{Bulk viscous corrections to
  screening and damping in QCD at high temperatures}},
  \href{http://dx.doi.org/10.1007/JHEP01(2017)123}{\emph{JHEP} {\bfseries 01}
  (2017) 123}, [\href{https://arxiv.org/abs/1611.08379}{{\ttfamily
  1611.08379}}].

\bibitem{Biondini:2017qjh}
S.~Biondini, N.~Brambilla, M.~A. Escobedo and A.~Vairo, \emph{{Momentum
  anisotropy effects for quarkonium in a weakly-coupled quark-gluon plasma
  below the melting temperature}},
  \href{http://dx.doi.org/10.1103/PhysRevD.95.074016}{\emph{Phys. Rev.}
  {\bfseries D95} (2017) 074016},
  [\href{https://arxiv.org/abs/1701.06956}{{\ttfamily 1701.06956}}].

\bibitem{Strickland:2011mw}
M.~Strickland, \emph{{Thermal $\Upsilon_{1s}$ and $\chi_{b1}$ suppression in
  $\sqrt{s_{NN}}=2.76$ TeV Pb-Pb collisions at the LHC}},
  \href{http://dx.doi.org/10.1103/PhysRevLett.107.132301}{\emph{Phys. Rev.
  Lett.} {\bfseries 107} (2011) 132301},
  [\href{https://arxiv.org/abs/1106.2571}{{\ttfamily 1106.2571}}].

\bibitem{Strickland:2011aa}
M.~Strickland and D.~Bazow, \emph{{Thermal Bottomonium Suppression at RHIC and
  LHC}}, \href{http://dx.doi.org/10.1016/j.nuclphysa.2012.02.003}{\emph{Nucl.
  Phys.} {\bfseries A879} (2012) 25--58},
  [\href{https://arxiv.org/abs/1112.2761}{{\ttfamily 1112.2761}}].

\bibitem{Krouppa:2015yoa}
B.~Krouppa, R.~Ryblewski and M.~Strickland, \emph{{Bottomonia suppression in
  2.76 TeV Pb-Pb collisions}},
  \href{http://dx.doi.org/10.1103/PhysRevC.92.061901}{\emph{Phys. Rev.}
  {\bfseries C92} (2015) 061901},
  [\href{https://arxiv.org/abs/1507.03951}{{\ttfamily 1507.03951}}].

\bibitem{Krouppa:2016jcl}
B.~Krouppa and M.~Strickland, \emph{{Predictions for bottomonia suppression in
  5.023 TeV Pb-Pb collisions}},
  \href{http://dx.doi.org/10.3390/universe2030016}{\emph{Universe} {\bfseries
  2} (2016) 16}, [\href{https://arxiv.org/abs/1605.03561}{{\ttfamily
  1605.03561}}].

\bibitem{Krouppa:2017lsw}
B.~Krouppa, R.~Ryblewski and M.~Strickland, \emph{{Bottomonia suppression in
  heavy-ion collisions}},  in \emph{{26th International Conference on
  Ultrarelativistic Nucleus-Nucleus Collisions (Quark Matter 2017)
  Chicago,Illinois, USA, February 6-11, 2017}}, 2017.
\newblock \href{https://arxiv.org/abs/1704.02361}{{\ttfamily 1704.02361}}.

\bibitem{Carrington:1997sq}
M.~E. Carrington, D.-f. Hou and M.~H. Thoma, \emph{{Equilibrium and
  nonequilibrium hard thermal loop resummation in the real time formalism}},
  \href{http://dx.doi.org/10.1007/s100520050412,
  10.1007/s100529800996}{\emph{Eur. Phys. J.} {\bfseries C7} (1999) 347--354},
  [\href{https://arxiv.org/abs/hep-ph/9708363}{{\ttfamily hep-ph/9708363}}].

\bibitem{Carrington:1998jj}
M.~E. Carrington, D.-f. Hou and M.~H. Thoma, \emph{{Ward identities in
  nonequilibrium QED}},
  \href{http://dx.doi.org/10.1103/PhysRevD.58.085025}{\emph{Phys. Rev.}
  {\bfseries D58} (1998) 085025},
  [\href{https://arxiv.org/abs/hep-th/9801103}{{\ttfamily hep-th/9801103}}].

\bibitem{Mrowczynski:2000ed}
S.~Mrowczynski and M.~H. Thoma, \emph{{Hard loop approach to anisotropic
  systems}}, \href{http://dx.doi.org/10.1103/PhysRevD.62.036011}{\emph{Phys.
  Rev.} {\bfseries D62} (2000) 036011},
  [\href{https://arxiv.org/abs/hep-ph/0001164}{{\ttfamily hep-ph/0001164}}].

\bibitem{Mrowczynski:2016etf}
S.~Mrowczynski, B.~Schenke and M.~Strickland, \emph{{Color Instabilities in the
  Quark-Gluon Plasma}},  \href{https://arxiv.org/abs/1603.08946}{{\ttfamily
  1603.08946}}.

\bibitem{Laine:2006ns}
M.~Laine, O.~Philipsen, P.~Romatschke and M.~Tassler, \emph{{Real-time static
  potential in hot QCD}},
  \href{http://dx.doi.org/10.1088/1126-6708/2007/03/054}{\emph{JHEP} {\bfseries
  03} (2007) 054}, [\href{https://arxiv.org/abs/hep-ph/0611300}{{\ttfamily
  hep-ph/0611300}}].

\bibitem{Arnold:2005ef}
P.~B. Arnold and G.~D. Moore, \emph{{QCD plasma instabilities: The NonAbelian
  cascade}}, \href{http://dx.doi.org/10.1103/PhysRevD.73.025006}{\emph{Phys.
  Rev.} {\bfseries D73} (2006) 025006},
  [\href{https://arxiv.org/abs/hep-ph/0509206}{{\ttfamily hep-ph/0509206}}].

\bibitem{Arnold:2005qs}
P.~B. Arnold and G.~D. Moore, \emph{{The Turbulent spectrum created by
  non-Abelian plasma instabilities}},
  \href{http://dx.doi.org/10.1103/PhysRevD.73.025013}{\emph{Phys. Rev.}
  {\bfseries D73} (2006) 025013},
  [\href{https://arxiv.org/abs/hep-ph/0509226}{{\ttfamily hep-ph/0509226}}].

\bibitem{Rebhan:2005re}
A.~Rebhan, P.~Romatschke and M.~Strickland, \emph{{Dynamics of
  quark-gluon-plasma instabilities in discretized hard-loop approximation}},
  \href{http://dx.doi.org/10.1088/1126-6708/2005/09/041}{\emph{JHEP} {\bfseries
  09} (2005) 041}, [\href{https://arxiv.org/abs/hep-ph/0505261}{{\ttfamily
  hep-ph/0505261}}].

\bibitem{Strickland:2007fm}
M.~Strickland, \emph{{Thermalization and the chromo-Weibel instability}},
  \href{http://dx.doi.org/10.1088/0954-3899/34/8/S31}{\emph{J. Phys.}
  {\bfseries G34} (2007) S429--436},
  [\href{https://arxiv.org/abs/hep-ph/0701238}{{\ttfamily hep-ph/0701238}}].

\bibitem{Dumitru:2006pz}
A.~Dumitru, Y.~Nara and M.~Strickland, \emph{{Ultraviolet avalanche in
  anisotropic non-Abelian plasmas}},
  \href{http://dx.doi.org/10.1103/PhysRevD.75.025016}{\emph{Phys. Rev.}
  {\bfseries D75} (2007) 025016},
  [\href{https://arxiv.org/abs/hep-ph/0604149}{{\ttfamily hep-ph/0604149}}].

\bibitem{Rebhan:2008uj}
A.~Rebhan, M.~Strickland and M.~Attems, \emph{{Instabilities of an
  anisotropically expanding non-Abelian plasma: 1D+3V discretized hard-loop
  simulations}},
  \href{http://dx.doi.org/10.1103/PhysRevD.78.045023}{\emph{Phys. Rev.}
  {\bfseries D78} (2008) 045023},
  [\href{https://arxiv.org/abs/0802.1714}{{\ttfamily 0802.1714}}].

\bibitem{Attems:2012js}
M.~Attems, A.~Rebhan and M.~Strickland, \emph{{Instabilities of an
  anisotropically expanding non-Abelian plasma: 3D+3V discretized hard-loop
  simulations}},
  \href{http://dx.doi.org/10.1103/PhysRevD.87.025010}{\emph{Phys. Rev.}
  {\bfseries D87} (2013) 025010},
  [\href{https://arxiv.org/abs/1207.5795}{{\ttfamily 1207.5795}}].

\bibitem{Laine:2007qy}
M.~Laine, O.~Philipsen and M.~Tassler, \emph{{Thermal imaginary part of a
  real-time static potential from classical lattice gauge theory simulations}},
  \href{http://dx.doi.org/10.1088/1126-6708/2007/09/066}{\emph{JHEP} {\bfseries
  09} (2007) 066}, [\href{https://arxiv.org/abs/0707.2458}{{\ttfamily
  0707.2458}}].

\bibitem{Schenke:2006xu}
B.~Schenke, M.~Strickland, C.~Greiner and M.~H. Thoma, \emph{{A Model of the
  effect of collisions on QCD plasma instabilities}},
  \href{http://dx.doi.org/10.1103/PhysRevD.73.125004}{\emph{Phys. Rev.}
  {\bfseries D73} (2006) 125004},
  [\href{https://arxiv.org/abs/hep-ph/0603029}{{\ttfamily hep-ph/0603029}}].

\end{thebibliography}\endgroup

\end{document}